\documentclass[%
 aip,
 jmp,%
 amsmath,amssymb,
 reprint,%
]{revtex4-1}

\usepackage{dcolumn}
\usepackage{bm}
\usepackage{lineno}

\providecommand{\Lt}{{\tt L}}
\renewcommand{\Lt}{{\tt L}}
\providecommand{\Wt}{{\tt W}}
\renewcommand{\Wt}{{\tt W}}

\def\bR{{\mathbb{R}}}

\def\e{\epsilon}

\def\t{\tau}

\def\cA{{\cal A}}

\def\cC{{\cal C}}

\def\cL{{\cal L}}

\def\cN{{\cal N}}

\def\cQ{{\cal Q}}

\def\cS{{\cal S}}

\def\cW{{\cal W}}

\def\cZ{{\cal Z}}

\def\be{\begin{equation}}
\def\ee{\end{equation}}
\def\bea{\begin{eqnarray}}
\def\eea{\end{eqnarray}}
\def\ba{\begin{array}}
\def\ea{\end{array}}
\def\nn{\nonumber}

\begin{document}

\preprint{AIP/123-QED}

\title{On the Explicit Asymptotic $\mathcal{W}_5$  Symmetry of 3D \\ Chern-Simons  Higher Spin $AdS_3$ Gravity}

\author{H. T. \"Ozer}
\email{ozert@itu.edu.tr}
\author{Ayt\"ul Filiz}%
\email{aytulfiliz@itu.edu.tr}
\affiliation{
\it{Istanbul Technical University,\,Faculty of Science and Letters,\,Physics Department,\,34469 Maslak,\,Istanbul,Turkey}
}%


\begin{abstract}
In this paper,\,we explicitly construct an asymptotic $\mathcal{W}_5$ symmetry algebra of the three-dimensional anti-de Sitter $(AdS_3)$ higher spin gravity.\,We use an $\mathfrak{sl}(5,\bR) \oplus \mathfrak{sl}(5,\bR)$  Lie algebra valued Chern-Simons gauge theory with a negative cosmological constant and its asymptotic symmetry algebra is explicitly calculated as two copies of the classical $\mathcal{W}_5$ algebra with central charge c.\,Our results can be interpreted as a spin\,$5$ extension of $AdS_3$ gravity and a proof of how the higher spin Ward identities and as well as the asymptotic $\mathcal{W}_5$  symmetry algebra are derived from the higher spin bulk field equations of motion.\,This higher spin asymptotic $\mathcal{W}_5$  symmetry algebra contains a finite number of conformal primary spin s:\,$s\,=\,2,\,3,\,4,\,5.$\,We also indicated how to introduce chemical potentials and holonomy conditions associated with these higher spin charges in $AdS_3$ higher spin gravity in a manner that it preserves the asymptotic symmetry algebra.
%
\end{abstract}

\maketitle
\section{Introduction}
Holographic dualities,\,which have found many remarkable applications that range from mathematics to modern theoretical physics during the last twenty years,\,claim an equivalence between gravitational theories defined in the bulk of some region of space-time and field theories on the boundary of that region.\,By now it is studied in many diverse subfields,\,and the literature on the subject has become enormous.\,The most famous realization of the holographic duality is the $AdS/CFT$ correspondence that was first conjectured by Juan Maldacena \cite{Maldacena:1997re} in 1997.\,To put it simply,\,the correspondence says that a theory of quantum gravity with a negative cosmological constant defined in $AdS_5$ is equivalent to a certain $CFT_4$ living on its boundary.

In the case of three dimensions,\,the space $AdS_3$ has the two-dimensional conformal group $SO(2,2)$ as isometry group.\,This group acts on the two-dimensional boundary of $AdS_3$ as the symmetry of the two-dimensional conformal field theory $CFT_2$.\,It is well-known that  $CFT_2$  has a fundamental role in modern physics.\,Therefore it has always been  of  interest to see to what degree a Virasoro algebra,\,which is its underlying symmetry algebra as the local extension of the global conformal algebra $SO(2,2)$,\,may be extended.\,In this context,\,$\cW_N$\, algebras are extensions of the Virasoro algebra by currents of higher spin. \cite{Zam85,Fat87,Luk87,Luk90,Fat90}\,These algebras are available for arbitrary values of the central charge\,$c$.

Another intriguing $\cW$\,algebra which is very closely linked to $\cW_{\it{N}}$\,algebra is $\cW_{\infty}$ family.\,$\cW_N$\,algebra can then be commented in the context of a $\cW_{\infty}$\,algebra.\cite{Pope:1989ew,Pope:1989sr,Pope:1990kc,Lu:1991pe}\,Such an algebra has an infinite set of higher spin,\,$s\geq 1$ generators.\,It should be emphasized  that the $\cW_{\infty}$ has been studied for special values of parameters where the algebra  linearizes.\,Besides,\,$\cW_{\it{N}}$\, algebras have been studied also for generic values of the parameters,\,see e.g.\,the review by Gaberdiel and Gopakumar.\,\cite{Gaberdiel:2012uj}\,The first marks of the connection between $\cW_{\infty}$ and $\cW_{\it{N}}$ is given in the pioneering work.\,\cite{Lu:1991pe}\,Based on the contemporary point of view,\,one can say that $\cW_{\infty}$ is an algebra with two parameters.\,\cite{Gaberdiel:2011wb,Gaberdiel:2011zw,Gaberdiel:2012ku}\,The connection between $\cW_{\it{N}}$ algebras with one parameter,\,particularly central charge $c$,\,and $\cW_{\infty}$ algebra family with two-parameters is analogous to the construction of the higher spin algebra $\mathfrak{hs}(\lambda)$ in Vasiliev theory \cite{Vasiliev:1999ba} and its universal relation to $\mathfrak{sl}(N)$ algebras.\,In an exactly related context of the extended  $\mathcal{W}$ algebras,\,one can say that the $AdS_3$ higher spin gravity\cite{Frad86,Frad87,Vasil92} is an interesting extension of the pure Einstein gravity.\,Einstein gravity theory is  a lot easier to handle  in three dimensions than in higher dimensions,\,because it allows a reformulation in terms of a Chern-Simons gauge theory.\,\cite{Achucarro:1987vz,Witten:1988hc}\,It has also been shown by Brown and Henneaux in their seminal work \cite{Brown:1986nw} that the asymptotic symmetry of $AdS_3$ gravity is given by two copies of a classical Virasoro algebra.\,This result can be interpreted as a pioneer application of the $AdS/CFT$ correspondence.\,Then application was generalized by Henneaux and Rey \cite{Henneaux:2010xg} and Campoleoni et al. \cite{Campoleoni:2010zq,Campoleoni:2011hg}[the authors provide a closed formula for the structure constants of all classical $W_{N}$ algebras where the approach depends on obtaining the algebras in the Poisson bracket notation.\,Our equivalent approach is the Operator Product Expansion (OPE) notation]to higher spin extensions on $AdS_3$.

Another route for the asymptotic symmetry algebra,\,instead of classical one,\,is the quantum $\mathcal{W}_N$ minimal models.\,\cite{Belavin:1984vu}\,The quantum asymptotic symmetry algebra of the $AdS_3$ higher spin gravity can be seen in the context of the minimal model holographic dualities.\,In particular,\,the $\mathcal{W}_N$ minimal models,\,which describe a family of two-dimensional $CFT_2$ with finite values of parameter $N$ and the central charge $c$ are dual to  higher spin theories in $AdS_3$.\,\cite{,Gaberdiel:2010pz}\,More importantly,\,the quantization problems are partially solved by Gaberdial and Gopakumar in Ref.\,\onlinecite{Gaberdiel:2012ku}.\,In this paper,\,the quantum corrections to the classical $\cW_{\infty}$ algebra have been determined by using the Jacobi identity as well as the representation theory of the $\cW_{\it{N}}$  minimal models.\,It is very interesting to calculate related quantum corrections directly in the higher spin gravity theory.

The impressive progress in the classification of the asymptotic symmetry algebras in recent  years indicates that it is necessary to describe the $AdS_3$ higher spin  gravity  theories  for sufficiently large conformal spin values.\,Therefore,\,the higher spin extension of the $AdS_3$  higher spin  gravity theories is still  an open problem because it is not yet known at least whether to be consistent with both small and large central charges.\,The reason for this is fundamentally the appearance of the composite or nonlinear terms in the related asymptotic symmetry algebras.\,We emphasize here that the appearance of these composite terms in the related asymptotic symmetry algebras have caused some serious problems from the proper quantum field theory point of view.\,That is,\,they mean that  the semi-classical  asymptotic symmetry  algebra must be corrected in order to guarantee that the Jacobi identities hold.\,To this end, one straightforward way is to replace the usual semi\,-\,classical asymptotic symmetry algebra with a quantum version one.\cite{Henneaux:2010xg}\,Finally,\,a deformed version of usual semi-classical asymptotic symmetry algebra with new structure constants can be obtained.\,This will be the final form of the quantum asymptotic symmetry algebra of the $AdS_3$ higher spin gravity.\,\cite{Zam85,Mathieu:1988pm,Bakas:1989mx,Ozer:2002xe}

The higher spin gravity theories have also come to play recently for their remarkable role in the context of $AdS/CFT$ correspondence.\,In three dimensions,\,within the setting  of $AdS_3/CFT_2$ duality,\,the difficulty of the higher spin gravity theories can be simplified as it is possible to truncate\,\cite{Campoleoni:2010zq} from an infinite number of spin fields to a set of the finite spin fields in the  $\mathfrak{sl}(N,\bR) \oplus \mathfrak{sl}(N,\bR)$ Chern-Simons theory.\,In this convention,the $\mathfrak{sl}(N,\bR)$ Lie algebra  belongs to ${SL}(N,\bR)$ group.\,Specifically,\,in the $\mathfrak{sl}(5,\bR) \oplus \mathfrak{sl}(5,\bR)$ Chern\,-\,Simons theory,\,the higher spin $AdS_3$ gravity  has asymptotic classical $\mathcal{W}_5$ symmetry algebra,\,which is the focus of the present paper.\,We emphasize  that the main idea of this type of works is that of Hamiltonian reduction through AdS\,-\,type boundary conditions,\,therefore $\mathcal{W}_N$ symmetry algebras should follow straightforwardly for sufficiently large values of $N$,\,in the  $\mathfrak{sl}(N,\bR) \oplus \mathfrak{sl}(N,\bR)$ Chern\,-\,Simons theory.\,Therefore,\,this work is also concerned with two-dimensional $CFT_2$ that appears in the holographic duality with higher spin theory on three dimensional $AdS_3$ space.\,The $CFT_2$ can be formulated as asymptotic $\mathcal{W}_5$ symmetry algebra which is higher-spin generalisation of the Virasoro algebra describing conformal symmetry with higher spin charges.

Since the asymptotic conditions allow incorporating the chemical potentials conjugated to the higher spin charges,\,one has  the chance to identify the thermodynamic properties of the higher spin gravity.\,A detailed recent discussion of higher spin black holes with chemical potentials are claimed and fulfilled in Ref.\,\onlinecite{Henneaux:2013dra} and also proposed independently in the form of new black hole solutions with spin 3 fields  by Gutperle and Kraus\,in the pioneering paper.\,\cite{Gutperle:2011kf}\,There and in following works involving the same authors the case of spin 3 black holes has been analysed in great detail,\,while generic higher spin black holes have been discussed only marginally.\,This led to succesive papers focussing on the spin 4 case\,(see e.g.\,Refs.\,\onlinecite{Tan,Chen1,Chen2,Ferlaino1,Beccaria,deBoer}),\,aiming at finding further support for the proposal of Ref.\,\onlinecite{Gutperle:2011kf} in a wider context.\,Especially,\,one can note that the case of spin 4 black hole story has been generalized without complication to the $\mathfrak{sl}(4,\bR) \oplus \mathfrak{sl}(4,\bR)$ case. \cite{Tan}\,In the gauge theories chemical potentials are presented by adding the time-component of the gauge connection.\,Therefore,\,it is also possible to carry out these studies entirely by adding  chemical potentials  associated to higher spin charges.\,It is well known that the chemical potential associated to spin 2 charge in pure gravity defines the temperature of the black hole.\,The procedure of adding a chemical potential to the connection in the higher spin algebra $\mathfrak{sl}(3,\bR)$ gravity is discussed \cite{Henneaux:2013dra,Bunster:2014mua} as an additional contributions to the thermal circles around the horizon of the black hole,\,and also showed that introducing of the chemical potential does not modify the asymptotic $\mathcal{W}_3$  symmetry algebra.\,Besides that,\,in the seminal work of Gutperle and Kraus \cite{Gutperle:2011kf} it was proposed that a higher spin black hole should be defined as a flat connection with a trivial holonomy along the thermal Euclidean cycle on the torus.\,The holonomy along the non-contractible spatial cycle on the other hand,\,was proposed to define different black hole solutions.\,We also emphasize here that in the literature on higher-spin black holes there are a few quantities that have been evaluated in the $\mathfrak{sl}(3,\bR)$ as well as in the $\mathfrak{sl}(4,\bR)$ case,\,but whose extension to the $\mathfrak{sl}(N,\bR)$ Chern-Simons theories has been only sketched.

Motivated by these developments in three-dimensional anti-de Sitter $(AdS_3)$ higher spin gravity we construct a classical solution based on an $\mathfrak{sl}(5,\bR) \oplus \mathfrak{sl}(5,\bR)$
 Lie algebra valued Chern-Simons gauge theory with a negative cosmological constant.\,Using some technique  recently found,\,we calculated explicitly a solution that can be interpreted as spin-5
 generalization of Ba$\tilde{n}$ados–Teitelboim–Zanelli (BTZ) solution.\,That is,\,we have constructed a spin-5 extension of Einstein gravity in three dimensions.

This paper has the following outline.\,In Sec.\,\ref{sec2},\,we give a fundamental formulation of pure gravity within $\mathfrak{sl}(2,\bR) \oplus \mathfrak{sl}(2,\bR)$ Chern\,-\,Simons theory in three-dimensions.\,Sec.\,\ref{hscs} is particularly devoted to the case of spin 5 where we showed in this section the principal embedding of $\mathfrak{sl}(2,\bR) $ in $\mathfrak{sl}(5,\bR)$,\,and also demonstrated how $\mathcal{W}_5$ symmetry and  higher spin Ward identities arise from the bulk equations of motion coupled to spin $s$,\,($s\,=\,3,4,5$) currents.\,Finally,\,classical $\mathcal{W}_5$ symmetry algebra as asymptotic spin 5 symmetry algebra, the chemical potentials appearing through the temporal components of the connection and holonomy conditions are obtained,\,and also the results are checked with the quantum $\mathcal{W}_5$  algebra.\,We conclude with a summary for our results and a few suggestions for future work.
\,Appendix \ref{appA},\,\ref{appB} and \ref{appC} collect our conventions for the  $\mathfrak{sl}(2,\bR)$,\,$\mathfrak{sl}(3,\bR)$ and $\mathfrak{sl}(5,\bR)$ Lie algebra generators respectively.\,Appendix \ref{hol33} also contains  holonomies for $\mathfrak{sl}(3,\bR) \oplus \mathfrak{sl}(3,\bR)$ Chern\,-\,Simons Theory.\,Conformal spin 5 Ward identities are introduced in Appendix\,\ref{appD}.\,The classical $\mathcal{W}_5$ symmetry algebra is presented in Appendix \ref{appE}.\,Finally,\,Appendix \ref{cp5} contains the chemical potentials for $\mathfrak{sl}(5,\bR) \oplus \mathfrak{sl}(5,\bR)$ Chern\,-\,Simons Theory.

\section{Gravity in Three Dimensions,\,a review:}\label{sec2}
Three dimensions is a good candidate because of its topological nature,\,which is a consequence of the lack of degrees of freedom.\,Chern-Simons theory is a quantum theory in three dimensions that computes only topological invariants.\,It can be defined on any manifold,\,and the metric does not need to be specified as it is a topological theory.\,Thus the physical quantities do not depend on the local geometry.\,Chern-Simons gravity is also a gauge theory and  used as an interesting playground for investigating the $AdS_3/CFT_2$ correspondence by Brown and Henneaux in their seminal work \cite{Brown:1986nw} in 1986.\,They showed that any quantum gravity theory with asymptotically $AdS_3$ boundary conditions in three-dimensions must be dual to $CFT_2$ at the meaning that asymptotic symmetries of $AdS_3$ are given by two duplicate copies of the Virasoro algebra and related the central charge $c$ of the $CFT_2$ to the corresponding $AdS_3$ radius $\ell$  in the following way
\begin{equation}\label{central}
c\,=\,\frac{3\ell}{2G}.
\end{equation}
This Brown Henneaux$\,-\,$formula,\,as one entry in the $AdS_3/CFT_2$ dictionary,\,notes that the central charge on the left-hand side is defined in the $CFT_2$,\,while the quantities at the right-hand side are defined in $AdS_3$.
\subsection{Connection to Chern\,-\,Simons Theory}\label{ccs}
It is a striking fact that the vacuum Einstein $AdS_3$ gravity in three dimensions with a negative cosmological constant can be formulated as a Chern\,-\,Simons gauge theory,\,as it was first proposed by Achucarro and Townsend in Ref.\,\onlinecite{Achucarro:1987vz} and developed by Witten in Ref.\,\onlinecite{Witten:1988hc}.\,One can start by defining  1\,-\,forms $(\mathcal{A},\bar{\mathcal{A}})$ taking values in the gauge group's $\mathfrak{sl}(2 ,\bR)$  Lie algebra,\,and the trace is taken over the algebra generators.\,The Chern\,-\,Simons action can be written in the form,
\begin{equation}
S = S_{CS}[\mathcal{A}] - S_{CS}[\bar{\mathcal{A}}]\\
\end{equation}
where
\begin{equation}
S_{CS}[\mathcal{A}] = \frac{k}{4\pi}\int \mathfrak{tr}\bigg( \mathcal{A}\wedge \mathrm{d}\mathcal{A}\,+\,\frac{2}{3}\,\mathcal{A}\wedge\mathcal{A}\wedge\mathcal{A} \bigg).
\end{equation}
Here $k=\frac{\ell}{8G\mathfrak{tr}( {\tt L}_{0} {\tt L}_{0}) }=\frac{c}{12\mathfrak{tr}( {\tt L}_{0} {\tt L}_{0}) }$ is the level of the Chern-Simons theory depending on the $AdS$ radius $l$ and the Newton's constant $G$  with the related  central charge $c$ of the $CFT_2$.\,Nevertheless,\,trace shows a metric on the $\mathfrak{sl}(2 ,\bR) $ Lie algebra.\,If $\Lt_{a},(a=\pm1,0)$ are the generators of $\mathfrak{sl}(2 ,\bR)$  Lie algebra,
\begin{equation}
[ {\tt L}_{a}, {\tt L}_{b}]=(a-b){\tt L}_{a+b}
\end{equation}
one can define  the invariant bilinear form
\begin{equation}
\mathfrak{tr}( {\tt L}_{a} {\tt L}_{b})  = \frac{1}{2}\eta_{ab}.
\end{equation}
For a negative cosmological constant $\mathcal{A}$ and $\bar {\mathcal{A}}$ are $\mathfrak{sl}(2 ,\bR) $ Lie algebra valued one-form,\,and they depend on the veilbein $e_{\mu}^a$ and dual spin connection $\omega^{a}_{\mu} = \e^{abc}\omega_{bc \mu}$ as follows
\begin{equation}\label{connections}
\mathcal{A} = \left(\omega^{a}_{\mu} + \frac{e^{a}_{\mu}}{\ell} \right) {\tt L}_{a} dx^{\mu}, \quad \bar{\mathcal{A}} = \left(\omega^{a}_{\mu} - \frac{e^{a}_{\mu}}{\ell} \right) {\tt L}_{a} dx^{\mu}.
\end{equation}
The equations of motion for the Chern\,-\,Simons gauge theory give the flatness condition $F = \bar{F} =0$ where
\begin{equation}
\label{flatness}
F= \mathrm{d}\mathcal{A} + \mathcal{A}\wedge \mathcal{A} =0,
\end{equation}
is the same as the Einstein's equation.\,$\mathcal{A}$ and $\bar{\mathcal{A}}$ are related to the metric $g_{\mu \nu}$ through the veilbein $e = \frac{\ell}{2}(\mathcal{A}- \bar{\mathcal{A}})$
\begin{equation}\label{metric}
g_{\mu \nu} =  \frac{1}{2}\,\mathfrak{tr}( e_{\mu} e_{\nu} )
\end{equation}
\subsection{$\mathfrak{sl}(2,\bR) \oplus \mathfrak{sl}(2,\bR)$ Chern\,-\,Simons Theory }\label{sl2}
Spin 2 case is given by reviewing asymptotically $AdS_3$ boundary conditions  for a $\mathfrak{sl}(2,\bR) \oplus \mathfrak{sl}(2,\bR)$ Chern\,-\,Simons theory,\,and how to determine the asymptotic symmetry algebra using the methods described in Ref.\,\onlinecite{Coussaert:1995zp}.\,By using the Fefferman\,-\,Graham expansion method,\,the most general solution of the Einstein's equation  that is asymptotically $AdS$,\,is given by with a flat boundary metric \cite{Banados:1998gg}
\begin{equation} \label{sol_einstein}
ds^2 = \, l^2 \left\{\, d\rho^2 - \frac{8\pi G}{l} \left(\cL\, (dx^+)^2 + \widetilde{\cL}\, (dx^-)^2\right) - \left(\, e^{2\rho} + \frac{64\pi^2G^2}{l^2}\,\cL\,\widetilde{\cL}\, e^{-2\rho}\right) dx^+ dx^-\right\}
\end{equation}
where $(\rho,x^{\pm}\equiv \frac{t}{\ell}\pm \phi)$ shows the solid cylinder as the light\,-\,like coordinates and $\cL\equiv\cL\,(x^+)$,$\widetilde{\cL}\equiv\widetilde{\cL}\,(x^-)$ are arbitrary functions
of $x^{\pm}$.\,Therefore,\,one can write the light\,-\,like components of the gauge fields by using the $\mathfrak{sl}(2,\bR)$ Lie algebra generators
\begin{eqnarray}
\label{ads31}
\mathcal{A}&=& b^{-1} a\left(x^{+} \right) b +b^{-1} db,\,\,\, \bar{\mathcal{A}}=b\bar{a}\left(x^-\right)b^{-1} + bdb^{-1},
\end{eqnarray}
with $b=e^{\rho \Lt_{0}}$,\,we obtain
\begin{eqnarray}\label{ads32}
a&=&\left(  \Lt_{1} - \frac{2\pi}{k} \cL \Lt_{-1}  \right) dx^+,\,\,\,\bar{a}=-\left( \Lt_{-1} - \frac{2\pi}{k} \tilde{\cL} \Lt_{1} \right)dx^-.
\end{eqnarray}
A very important point in this  description is that this theory has to be asymptotically $AdS_3$,\,as required in Ref.\,\onlinecite{Campoleoni:2010zq}  the boundary conditions have to be defined in a similar  conditions
\begin{equation}
\label{bc}
\left( \mathcal{A} - \mathcal{A}_{AdS_3} \right)  \Big |_{\textrm{boundary}} = \mathcal{O}(1).
\end{equation}
We will carry out such an analysis for a connection of the spin 2 charge,\,finding that the asymptotic symmetry can be identified with a classical $\mathcal{W}_2$-algebra.\,After that,\,we will only work on the positive chiral components having $x^{+}$,\,although the same can be done in the other one as well.\,If we expand $\lambda(x^{+})$ in the $\mathfrak{sl}(2,\bR)$ Lie algebra,\,as we did also the connection,\,then
 \begin{equation}\label{lambda}
\lambda\,=\,\sum_{i=-1}^{1} \epsilon_i  \Lt_{i}.
\end{equation}
We are now interested in the transformation that preserve the structure of \eqref{ads31}.\,Under an infinitesimal gauge transformation with gauge parameter $\lambda$,\,$a$ which is equivalent to $\mathcal{A}$ transforms as the flatness condition:
\begin{equation}
\delta_\lambda a\,=\,d \lambda\,+\,[a,\lambda] \label{flat}.
\end{equation}
Thus we have to impose that all terms proportional to $\Lt_{0},\Lt_{1}$ vanish.\,These constraints can be solved to find $\epsilon_{0}$ and $\epsilon_{-1}$ in terms of $\epsilon_{1}$ and their
derivatives.\,Writing\,$\epsilon_{1}\equiv\epsilon$\,which is called the gauge parameter related to $\mathfrak{sl}(2,\bR)$ and superscripted primes denote $\partial_{x^+}$,\,one finds
\begin{align}
\epsilon_{0}&=-\epsilon', \nonumber \\
\epsilon_{-1}&=\frac{1}{2} \epsilon''-\frac{2 \pi}{k}\epsilon \mathcal{L}.
\end{align}
and \eqref{lambda} is of the form:
 \begin{equation}\label{lambda2}
\lambda(\epsilon)\,=\,\epsilon \Lt_{1}-\epsilon' \Lt_{0}+\frac{1}{2}\Big(\epsilon''-\frac{4 \pi}{k}\epsilon\mathcal{L}\Big)\Lt_{-1}.
\end{equation}
Now,\,one can also determine how the function $\mathcal{L}$  transforms under this gauge transformation.\,This is given by
 \begin{equation}
\mathcal{L} \rightarrow \mathcal{L}+\delta _{\epsilon }\mathcal{L},
\end{equation}
where
\begin{equation}\label{ward1}
\delta _{\epsilon }\mathcal{L} =2 \mathcal{L} \epsilon '+  \mathcal{L}'\epsilon+\frac{k}{4 \pi }\epsilon ^{'''}.
\end{equation}
As a final step,\,one now has to determine  the canonical boundary charge $\mathcal{Q}[\epsilon] $ that generates this transformation.\,Therefore,\,the corresponding variation of the boundary charge $\mathcal{Q}[\epsilon] $ can be integrated which reads
\begin{equation}\label{chargeQ}
\mathcal{Q}[\epsilon] \,=\,\int dx^{+}\epsilon(x^{+})\mathcal{L}(x^{+}).
\end{equation}
This leads to Dirac bracket algebra by using $\delta _{\epsilon }\digamma\,=\,\{\digamma,\mathcal{Q}[\epsilon]\}$
\begin{equation}\label{bracket}
\{\mathcal{L}(x^{+}),\mathcal{L}(y^{+})\}\,=\,2 \mathcal{L}(y^{+})\delta'(x^{+}-y^{+})+\mathcal{L}'(y^{+})\delta(x^{+}-y^{+})+ \frac{k}{4 \pi }\delta'''(x^{+}-y^{+}).
\end{equation}
One can also expand $\mathcal{L}(x^{+})$ into Fourier modes\,$\mathcal{L}(x^{+})=\sum_n{\Lt_{n}}e^{-inx^+}$,\,and replacing $i\{\cdot,\cdot\}\rightarrow[\cdot,\cdot]$.\,A Virasoro algebra can then be defined as:
\begin{equation}
[\Lt_{n},\Lt_{m}]=(n-m) \Lt_{n+m}\,+\, \frac{c}{12}n (n^2-1) \delta_{n+m,0}.
\end{equation}
We now consider the case that the conformal boundary is a complex cylinder,\,as in \eqref{sol_einstein}.\,We work in complex coordinates,\,$z(\bar{z})\equiv \phi \pm i\,\frac{t}{\ell}$.\,Therefore,\,the Virasoro algebra in these space is equivalent to operator product algebra
\begin{equation}\label{ope}
\mathcal{L}(z_1)\mathcal{L}(z_2)\,\sim \,{{c\over 2}\over{z_{12}^{4}}}\,+\, {2\,\mathcal{L}\over{z_{12}^{2}}}\, + \,{ \mathcal{L}'\over{z_{12}}}
\end{equation}
where the central charge $c$ is again \eqref{central} related to the level of the Chern-Simons theory as
\begin{equation}\label{central2}
c\,=\,6k\,=\,\frac{3\ell}{2G}.
\end{equation}
The same algebra is also realized as the asymptotic symmetry algebra of the anti-chiral connection $\bar{\mathcal{A}}$ having $x^{-}$ in terms of the second copy of $\mathfrak{sl}(2,\bR)$ Lie algebra.
\subsection{Adding a chemical potential to $\mathfrak{sl}(2,\bR) \oplus \mathfrak{sl}(2,\bR)$ Chern\,-\,Simons Theory }\label{cp}
\noindent The procedure of adding a chemical potential to the connection in pure $\mathfrak{sl}(2,\bR)$ gravity discussed in Ref.\,\onlinecite{Henneaux:2013dra} is an additional contribution to the thermal circles around the horizon of the black hole.\,One can show that an addition of the chemical potential to the connection does not modify the asymptotic $\mathcal{W}_2$  symmetry algebra as in (\ref{ward1})-(\ref{central2}).\,To do this,\,one can propose the following boundary conditions for the connection $a(t,\phi)$ in the Chern - Simons formulation,
\begin{eqnarray}
\label{ch33}
a(t,\phi)&=&a_{\phi}(t,\phi)d\phi +a_{t}(t,\phi)dt
\end{eqnarray}
where
\begin{equation}\label{spatial1}
a_{\phi}(t,\phi)\,=\, \Lt_{1} - \frac{2\pi}{k} \cL \Lt_{-1}
\end{equation}
and
\begin{equation}\label{temp}
a_{t}(t,\phi)\,=\,\mu_2 \Lt_{1}+ \sum_{i=-1}^{0} \nu_{2}^{(i)}  \Lt_{i}.
\end{equation}
Here $\mu_2$ is in principle arbitrary functions of $t$ and $\phi$.\,We interpret the function $\mu_2$ as a chemical potential.\,This means that we assume the chemical potential to be fixed at infinity,\,i.e.\,$\delta\mu_2=0$.\,The functions $\nu_{2}^{(0)}$ and $\nu_{2}^{(-1)}$ are fixed by the flatness condition (\ref{flatness}).\,For fixed chemical potential $\mu_2$,\,the time evolution of canonical boundary charge $\cL$ as well as $\nu_{2}^{(0)}$ and $\nu_{2}^{(-1)}$ can be written as
\begin{align}\label{evo}
\partial_{t }\mathcal{L}&=2 \mathcal{L} \mu_2 '+  \mathcal{L}'\mu_2+\frac{k}{4 \pi }\mu_2 ^{'''},\nonumber \\
\nu_{2}^{(0)}&=-\mu_2',\,\,\,
\nu_{2}^{(-1)}=\frac{1}{2} \mu_2''-\frac{2 \pi}{k}\mu_2 \mathcal{L}.
\end{align}
where prime denotes derivative with respect to $\phi$,\,and the temporal connection $a_{t}$ is of the form:
\begin{equation}\label{timecomp}
a_{t}\,=\,\mu_2\Lt_{1}-\mu_2' \Lt_{0}+\frac{1}{2}\Big(\mu_2''-\frac{4 \pi}{k}\mu_2\mathcal{L}\Big)\Lt_{-1}.
\end{equation}
In many applications,\,however,\,the chemical potential $\mu_2$ is constant,\,which can simplify most of the  formulas,\,especially in $AdS_3$ higher spin gravity,\,considerably.\,Therefore,\,one can note that from now on one will assume that the chemical potential $\mu_2$ as well as the canonical boundary charge $\cL$ are constant.\,Under this assumption,\,the spatial connection $a_{\phi}$ and the temporal connection $a_{t}$ simplify as
\begin{eqnarray}\label{comp12}
&a_{\phi}\,&=\, \Lt_{1} - \frac{2\pi}{k} \cL \Lt_{-1} \\
&a_{t}\,&=\,\mu_2\left( \Lt_{1} - \frac{2\pi}{k} \cL \Lt_{-1}\right)\label{comp13}
\end{eqnarray}
respectively.\,These manifestly solve the field equation(\ref{evo}),\,and describe black holes solutions carrying not only mass and angular momentum but also canonical charges of the pure gravity.\,As discussed in Ref.\,\onlinecite{Henneaux:2013dra},\,and comparison with e.g.\,eq.(2.29) of Ref.\,\onlinecite{Bunster:2014mua} shows that  the temporal component of the connection is linear in the chemical potential $\mu_2$.\,The same procedure can be followed to establish $\bar{a}$ with a constant chemical potential.
\subsection{Holonomies in  $\mathfrak{sl}(2,\bR) \oplus \mathfrak{sl}(2,\bR)$ Chern\,-\,Simons Theory }\label{hol0}

As we have seen,\,classical solutions of the  $\mathfrak{sl}(2,\bR) \oplus \mathfrak{sl}(2,\bR)$ Chern\,-\,Simons theory are given by the flat connections,\,and thus locally may be written as  pure gauges,\,i.e.\,$\mathcal{A} =g^{-1}\mathrm{d}g$.\,Globally,\,however this is not true as the space-time may have some non-trivial topology,\,then the gauge function $g$ is not single-valued.\,When the spacetime  has a non-contractable cycle $\mathcal{C}$,\,the gauge function $g$ attains a factor of holonomy $\mathcal{P}\mathrm{exp}(\oint_{\mathcal{C}}\mathcal{A_{\mu}\mathrm{d}^{\mu}})$.\,This means that classical solutions in Chern-Simons theory are uniquely specified by the holonomies,\,which are gauge invariant quantities,\,around the cycles of the manifold which is a solid torus,\,up to an overall gauge transformation.
\par The solutions of the Chern Simons theory are,\,in Euclidean signature,\,specified by two cycles:\,The spatial cycle and the thermal cycle.\,The thermal cycle is contractable,\,that is,\,the holonomy around the thermal cycle is trivial,\,which is usually related to the inverse temperature of the solution.\,But the $\phi$ coordinate represents the non-contractable cycle.\,Thus,\,the flat connection $a$'s are uniquely specified by the non-trivial holonomy:
\begin{equation}\label{hol1}
H_{\phi}(a)\,=\,\mathrm{b^{-1}}\mathrm{e}^w\mathrm{b},\,\,\,\,\, 0\,<\,\phi\,<2\,\pi
\end{equation}
where $w\,=\,\oint_{\mathcal{C}}a_{\phi}\mathrm{d}{\phi}\,=\,2\pi a_{\phi} $ is the holonomy matrix.\,For the Euclidean $BTZ$ black hole,\,the $\phi$-cycle is non-contractable,\,and will have a non-trivial holonomy matrix:
\begin{equation}\label{holmat2}
w\,=\,\,2\pi a_{\phi} \,=\,\begin{pmatrix} 0 &  -\frac{4\pi^2}{k} \cL\\ -2\pi & 0\end{pmatrix}.
\end{equation}
One must emphasize here that  what different values of $\cL$ give rise to different solutions.\,Besides,\,in many cases it might be cumbersome or even impossible to evaluate the holonomy matrix exactly,\,as was done
above.\,But often we do not need the full matrix and will be satisfied with knowledge of the eigenvalues of the exponentiated matrix $\mathrm{e}^w$.\,The easiest way to solve the holonomy condition is by solving the characteristic polynomial.\,This means that one may express exponent matrix $w$,\,a general $2\times2$ matrix,\,as
\begin{equation}\label{holmat3}
w^2\,=\,\Theta_0+\Theta_1 w
\end{equation}
where $\Theta$'s are referred to as the \textit{holonomy invariants}.\,To find this invariants one consider the general characteristic equation of exponent matrix $w$,\,explicitly
\begin{equation}\label{holmat4}
\lambda^2\,=\,4 \pi^2 \cQ_2=-(\lambda_1+\lambda_2)\lambda+\lambda_1\lambda_2
\end{equation}
where $\cQ_2\,=\,\frac{1}{2}\mathfrak{tr}[a_\phi^2]=\,\frac{2\pi}{k} \cL$ is a second order \textit{Casimir operator}-related parameter,\,and one can use the idendities
\begin{equation}\label{ident1}
\mathfrak{tr}[w]=\sum_i \lambda_i,\,\,\mathfrak{det}[w]=\prod_{i}\lambda_i,
\end{equation}
If $\lambda_i$ is $i^{th}$ eigenvalue of the exponentiated matrix $w$,\,and identify the \textit{holonomy invariants} in (\ref{holmat3}) as
\begin{equation}\label{holmat5}
\Theta_0=-\mathfrak{det}[w]= - 4\pi^2 \cQ_2,\,\,\,\,\Theta_1=\mathfrak{tr}[w]=0.
\end{equation}
therefore,\,since $\cL$ is non-negative,\,they are manifestly nontrivial.\,If a holonomy is trivial around a contractible cycle  the exponentiated matrix $\mathrm{e}^w$ must have eigenvalues $\pm \pi i$,\,it follows that
\begin{equation}\label{holmat8}
 \mathfrak{tr}[w]=0,\,\,\mathfrak{tr}[w^2]+2\pi^2=0.
\end{equation}
Now when it comes to calculating the holonomy about contractible (thermal) cycle one may use the temporal  component $a_t=\mu_2 a_\phi$ in (\ref{comp12}),\,which  fulfils the equation of motion $[a_t,a_\phi]=0$\,it is obvious that $a_t= \mathit{f}(a_\phi)$.\,Besides,\,it is important to recall that one usually works with a free periodicity on the temporal coordinate with Euclidean signature.\,This is usually done by first performing a Wick rotation ($t\rightarrow i \tau $) of the Lorentzian time coordinate t to Euclidean time $\tau$\,which is then compactified.\,This changes the topology of the solid cylinder to that of a solid torus.\,One can introduce complex coordinates $(z,\bar{z})$ by analytically continuing the light-cone directions $x^{\pm}\rightarrow i\frac{\tau}{\ell}\pm \phi\equiv(z,-\bar{z})$.\,Then $z\simeq z+2 \pi \tau$ where $\tau$ is the modular parameter of the boundary torus,\,which is usually related to the inverse temperature $\beta$ of the solution as $\tau=\frac{i\beta}{2\pi}$ and also compatible with chemical potential $\mu_2$ in (\ref{comp13}).\,Therefore,\,the trivial holonomy is
\begin{equation}\label{hol2}
H_{\t}(a)\,=\,\mathrm{b^{-1}}\mathrm{e}^w\mathrm{b},
\end{equation}
where $w\,=\,\oint_{\mathcal{C}}a_{t}\mathrm{d}{t}\,=\,2\pi\tau a_{t} $ is the holonomy matrix,\,and the eigenvalues of $w$ will be $\pm\pi i$ as non-degenerate for the connection considered here.\,Hence,\,we can use (\ref{holmat8}),\,it follows that
\begin{equation}\label{thermal0}
 \mathfrak{tr}[w]=0,\,\,\mathfrak{tr}[w^2]+2\pi^2=0.
\end{equation}
Finally,\,while not $ \mathbb{I}$,\,the exponentiated matrix  $e^w=-\mathbb{I}$ is in the center of $\mathfrak{sl}(2,\bR)$,\,so it is a trivial holonomy through the periodicity of $\beta$.\,In order to  solve the holonomy conditions,\,the entropy function can be written in the form
\begin{align}\label{entropy1}
\cS(\cL)\,=\,2\pi \sqrt{2\pi k \cL}
\end{align}
such that
\begin{align}\label{entropy2}
\tau =\,&i\frac{1}{4\pi^2 } \frac{\partial\cS}{\partial \cL}\,=\,\frac{ik}{2}\frac{1}{\sqrt{2\pi k \cL}}.
\end{align}
This solves the  holonomy condition,\,$\mathfrak{tr}[w^2]+2\pi^2=0$,\,with $\beta\,=\,\sqrt{\frac{\pi k}{2\cL}}$.

\section{Higher spin  Chern\,-\,Simons theory}\label{hscs}

The Chern-Simons formalism in three dimensions lends itself to generalization to higher spin theories coupled to gravity by raising the gauge group to $\mathfrak{sl}(N ,\bR) $.\,The $\mathfrak{sl}(N ,\bR)$ Chern-Simons theory describes $AdS_3$ gravity theories coupled to a finite tower of integer spin-s $\leq N$.\,Therefore,\,an interesting extension of pure $AdS$ gravity can be constructed by adding massless higher spin fields to the spectrum,\,which proves to be particularly easy in Chern\,-\,Simons formulation.\,The first nontrivial spin 3 extensions of higher spin theories coupled to the pure gravity can be seen in Refs.\,\onlinecite{Henneaux:2010xg,Campoleoni:2010zq},\,where a consistent set of asymptotic conditions for the theory is described.\,It can be shown that,\,instead of describing difficult nonlinear interactions of the higher spin fields,\,promoting the $\mathfrak{sl}(2,\bR) \oplus \mathfrak{sl}(2,\bR)$ symmetry group of the Chern\,-\,Simons action to $\mathfrak{sl}(N,\bR) \oplus \mathfrak{sl}(N,\bR)$,\,with $N > 2$,\,is sufficient to successfully describe a higher spin theory.

\subsection{A calculation of $\mathfrak{sl}(5,\bR)$ commutators  }\label{sl5com}
\label{sec:first-commutators}
The principal embedding of $\mathfrak{sl}(2,\bR)$ Lie algebra to $\mathfrak{sl}(5,\bR)$ Lie algebra contains  the spin 2 triplet $\Lt_{n},(n=\pm1,0)$,\,the spin 3 triplet $\Wt_{n}^{(3)},(n=\pm2,\pm1,0)$,\,the spin 4 quartet $\Wt_{n}^{(4)},(n=\pm3,\pm2,\pm1,0)$ and the spin 5 quintet $\Wt_{n}^{(5)},(n=\pm4,\pm3,\pm2,\pm1,0)$  can be represented in terms of the  commutation relations for $\Lt_{n}\equiv\Wt_{n}^{(2)}$.\\

Let us now explain how $\mathfrak{sl}(5,\bR)$ Lie algebra can be constructed.\,We proceed in two steps.\,First,\,we make the most general ansatz for the commutators between all the Lie algebra generators $\Wt_{n}^{(s_i)}$,\,$\Wt_{m}^{(s_j)}$ as follows
\begin{subequations}
  \begin{align}
    [\Lt_{n},\Wt_{m}^{(s_i)}]=((s_i-1)n-m) \Wt^{(s_i)}_{n+m}, (s_i=2,3,4,5)
      \\
    [\Wt_{n}^{(s_i)},\Wt_{m}^{(s_j)}]= \sum_{t}^{s_i+s_j}f_{s_i s_j}^{t}(n,m) \Wt_{n+m}^{(t)}.
       \end{align}
\end{subequations}
Our next goal is to find concrete expressions for the structure constants $f_{s_1s_2}^{s_3}(n,m)$.\,It can be seen here that some structure constants does not survive because of the spin gradation conformity.\,This structure constants which are the polynomials in $\it{n}$ and  $\it{m}$ indices are also standard and known in closed form and given as just the commutator structure of quasi primary field in a $CFT$,\,(see e.g.\,the book of Blumenhagen and Plauschinn \cite{Blumenhagen:2009zz} in Sec.\,2.6.3) as $f_{s_i s_j}^{s_k}(n,m)\,\sim\,p_{ijk}(n,m)$,\,where

\begin{subequations}
  \begin{align}
   & p_{ijk}(n,m)=\sum C_{r,s}^{ijk}\cdot\binom{-m+s_i-1}{r}\cdot\binom{-n+s_j-1}{s}
      \\
   &~~~~\begin{array}{c}{r,s}\in Z_0^{+}\\{r+s=s_i+s_j-s_k-1}\end{array}\nonumber
      \\
   & C_{r,s}^{ijk}=(-1)^r \frac{(2s_k-1)!}{(s_i+s_j+s_k-2)!}\prod_{t=0}^{s-1}(2s_i-2-r-t)\prod_{u=0}^{r-1}(2s_j-2-s-u).
       \end{align}
\end{subequations}
Then,\,we used the associativity of the $\mathfrak{sl}(5,\bR)$ Lie algebra with Jacobi identities under rescaling  $f_{s_i s_j}^{s_k}(n,m)=\sigma_a\,p_{ijk}(n,m)$.\,Since this  associativity condition is of central importance for this section,\,we will find 11 expressions for the structure constants $f_{s_1s_2}^{s_3}(n,m)$.\,(Further details of our conventions can be found in Appendix \ref{appC}).
\subsection{Spin 5 $AdS_3$ solution and conformal Ward identities}\label{spin5}

We now proceed to construct  $\mathfrak{sl}(5,\bR)$  connections which describe Spin 5 $AdS_3$ solution with non-zero higher spin charge.\,We consider the class of connection using the same gauge and notation as in Sec.\,\ref{sl2},\,one can write down the spin 5 extension of (\ref{ads32}) as Ref.\,\onlinecite{Campoleoni:2010zq}.\,In this part,\,we will now make a similar calculation for spin 5.\,We begin with the ansatz
\begin{eqnarray}
\label{ads322}
a&=&\left(  \Lt_{1} + \alpha \mathcal{L} \Lt_{-1}+ \beta \mathcal{W}_3 \Wt_{-2}^{(3)}+ \gamma \mathcal{W}_4 \Wt_{-3}^{(4)}+ \delta \mathcal{W}_5 \Wt_{-4}^{(5)}  \right)
     dx^+\label{a}\\
\bar a&=&-\left(  \Lt_{-1} + \alpha \tilde{\mathcal{L}} \Lt_{1}+ \beta \tilde{\mathcal{W}_3} \Wt_{2}^{(3)}+ \gamma \tilde{\mathcal{W}_4} \Wt_{3}^{(4)}+ \delta \tilde{\mathcal{W}_5} \Wt_{4}^{(5)}  \right) dx^-
\end{eqnarray}
where $\alpha,\beta,\gamma$ and $\delta$ are some scaling parameters to be determined later.\,After this,\,we analyze the holomorphic connection $a$.\,All resulting expressions involving the $\bar{A}$ sector can be obtained in the similar way.\,If we expand $\lambda(x^{+})$ in the $\mathfrak{sl}(5,\bR)$ Lie algebra,\,as we did also for the connection which is compatible with (\ref{a}),\,then
\begin{eqnarray}
\label{ads322}
\lambda\,=\,\sum_{i=-1}^{1} \epsilon_i  \Lt_{i} + \sum_{i=-2}^{2} \chi_i  \Wt_{i}^{(3)}  + \sum_{i=-3}^{3} \mathit{f}_i  \Wt_{i}^{(4)}
                 + \sum_{i=-4}^{4} \eta_i  \Wt_{i}^{(5)}.
\end{eqnarray}
The flat connection condition (\ref{flat}) emerges the following constraints for 24 undetermined functions of\,$x^{\pm}$.\,There are 24 equations with 4 parameters $\epsilon_1,\chi_2, \mathit{f}_3, \eta_4$ that can be chosen arbitrarily.\,For notational convenience,\,we write $\epsilon_1\equiv \epsilon,\chi_2\equiv \chi, \mathit{f}_3\equiv\mathit{f}, \eta_4\equiv\eta$,\,and show derivatives with respect to  ${x^+}$ as primes for convenience,\,one finds
\begin{subequations}
\begin{align}
\epsilon_{ 0}&\,=\,-\epsilon'\label{ep0}\\
\epsilon_{-1}&\,=\, \frac{69120 \gamma  \mathit{f} \sigma _ 1^2 \mathcal{W}_ 4}{49 \sigma _ 2^2}+24 \beta  \sigma _ 1 \chi  \mathcal{W}_ 3+\frac{3686400 \delta  \eta  \sigma _ 1^3 \mathcal{W}_ 5}{343 \sigma _ 2^2 \sigma _ 4^2}+\alpha  \mathcal{L} \epsilon +\frac{\epsilon ''}{2},
\end{align}
\end{subequations}
and all other parameters are given in Appendix \ref{par}.\,Then,\,one can now also determine how the functions $\mathcal{L}$,\,$\mathcal{W}_s$\,$(s\,=\,3,4,5)$  transform under the gauge transformation in terms of $\epsilon,\,\chi,\,\mathit{f},\,\eta,\,\mathcal{L}$,\,$\mathcal{W}_s$\,$(s\,=\,3,4,5)$ and their derivatives.\,The transformations of $\mathcal{L}$,\,$\mathcal{W}_s$\,$(s\,=\,3,4,5)$ are given by
\begin{subequations}
\begin{align}
\mathcal{L}& \rightarrow \mathcal{L}+\delta _{\epsilon }\mathcal{L}+\delta _{\chi}\mathcal{L},\\
\mathcal{W}_s& \rightarrow \mathcal{W}_s+\delta _{\epsilon}\mathcal{W}_s+\delta _{\chi}\mathcal{W}_s+\delta _{\mathit{f}}\mathcal{W}_s+\delta _{\eta}\mathcal{W}_s,\,(s\,=\,3,4,5)
\end{align}
\end{subequations}
where
\begin{equation}
\delta _{\epsilon }\mathcal{L} =2 \mathcal{L} \epsilon ' +  \mathcal{L}'\epsilon + \frac{\epsilon ^{'''}}{2 \alpha }
\end{equation}
here $\epsilon$ is the gauge parameter related to $\mathfrak{sl}(2,\bR)$ subgroup in $\mathfrak{sl}(5,\bR)$, which generate conformal transformations.\,Thus,\,we see that $\mathcal{L}$ can be identified with the $CFT_2$ stress tensor if we relate the Chern-Simons level and the central charge as $c=6k$.\,Now, first observe that if we set from $(\ref{ward1})$ and $(\ref{central2})$, we can write the first scaling parameter $\alpha$ as
\begin{equation}
\alpha\,=\,\frac{6}{c}
\end{equation}
which is compatible with the central charge of the Virasoro algebra $c\,=\, 6 k$.\,Finally we will set $\sigma_1\,=\,-\frac{1}{30},\,\sigma_2\,=\,\frac{16}{\sqrt{105}}$, $\sigma_4\,=\,\frac{\sqrt{15}}{7}$ , and also
$\beta\,=\,\frac{15}{c}\,,\,\gamma\,=\,\frac{28}{c}$ and $\delta\,=\,\frac{45}{c}$.\,In the light of these results,\,for the later calculations from now on we  generalize the connection $(\ref{a})$ for $\mathfrak{sl}(N,\bR)$ and  $ \Lt_{n}\equiv\Wt_{n}^{(2)}$  as
\begin{equation}
\label{a2}
a=\left(\Wt_{1}^{(2)} + \sum_{s=2}^{N}\frac{s (2s-1)}{c} \mathcal{W}_{s} \Wt_{-s+1}^{(s)}  \right) dx^+,
\end{equation}
in order to match $AdS_3$ connection parameters to the classical $\cW_5$ asymptotic symmetry algebra as in Appendix \,\ref{appE}.\,Then,\,the following variations
\begin{subequations}
\begin{align}
\delta _{\epsilon }\mathcal{W}_s&\,=\,s \mathcal{W}_s \epsilon '+\epsilon  \mathcal{W}_s'\,,(s=3,4,5)
\end{align}
\end{subequations}
define,\,that the each conformal field $ \mathcal{W}_s$ has the conformal spin s,\,with $s=3,4,5$.\,All variations which defines spin 5 conformal Ward idendities are given in Appendix \ref{Wi}.\,As in the spin 2 case in $(\ref{chargeQ})$,\,one can now determine the corresponding canonical boundary charges as
\begin{equation}\label{chargeQ2}
\mathcal{Q}[\epsilon] \,=\,\int dx^{+}\left( \epsilon(x^{+})\mathcal{L}(x^{+})+\chi(x^{+})\mathcal{W}_3(x^{+})+\mathit{f}(x^{+})\mathcal{W}_4(x^{+})+\eta(x^{+})\mathcal{W}_5(x^{+}) \right).
\end{equation}
In fact,\,from these Ward identities,\,that is,\,from the variation of these fields one can derive  their corresponding classical asymptotic $\mathcal{W}_5$ symmetry algebra as in Appendix \ref{appE}.
\subsection{Adding  chemical potentials to $\mathfrak{sl}(5,\bR) \oplus \mathfrak{sl}(5,\bR)$ Chern\,-\,Simons Theory }\label{sl5}

So far,\,we have not incorporated the chemical potentials in the $\mathfrak{sl}(5,\bR)$ Chern-Simons higher spin theory.\,We emphasize here that in the literature on higher-spin black holes there are a few quantities that have been evaluated in the $\mathfrak{sl}(N,\bR)$ case up to $N=4$,\,but whose extension to the $\mathfrak{sl}(N,\bR)$ Chern-Simons theories,\,$N>4$,\,has been only sketched.\,To do this,\,one must first concentrate on (\ref{ads322}) and twenty solved free parameters in Appendix \ref{par} depending on $ \epsilon,\chi,\mathit{f}$ and $\eta$. Here $\cL,\,\cW_i;\,(i=3,4,5)$ and  $ \epsilon,\chi,\mathit{f},\eta$ stand for arbitrary functions of  $(t,\varphi)$.\,The asymptotic behaviour of the spatial part of the connection is preserved under gauge transformations generated by $\lambda$ in\,(\ref{ads322}).

The procedure of adding a chemical potential to the connection in pure $\mathfrak{sl}(5,\bR)$ gravity discussed in Ref.\,\onlinecite{Henneaux:2013dra} is an additional contribution to the thermal circles around the horizon of the black hole.\,One can show that an addition of the chemical potential to the connection does not modify the asymptotic $\mathcal{W}_5$ symmetry algebra as in Sec.\,\ref{spin5}.\,Notice that the chemical potentials
$\mu_2,\mu_3,\mu_4$ and $\mu_5$ appear only in the components of the gauge fields  along $t$,\,and they are assumed to be fixed at infinity,\,$\delta\mu_2=\delta\mu_3=\delta\mu_4=\delta\mu_5=0$.\,As explained in  Refs.\,\onlinecite{Henneaux:2013dra} and \onlinecite{Bunster:2014mua},\,the chemical potentials are included in the temporal component of the gauge fields only,\,so that the asymptotic form of the the gauge fields with constant fields $\cL,\,\cW_i;\,(i=3,4,5)$ and chemical potentials $ \mu_2,\mu_3,\mu_4$ and $\mu_5$,\,provided that the spatial part of the connection is unchanged.\,To do this,\,one can propose the following boundary conditions for the connection $a(t,\phi)$ in the Chern - Simons formulation,
\begin{eqnarray}
\label{ch33}
a(t,\phi)&=&a_{\phi}(t,\phi)d\phi +a_{t}(t,\phi)dt
\end{eqnarray}
where
\begin{align}\label{ch34}
a_{\phi}(t,\phi)&=
  \Lt_{1}
+ \frac{6}{c} \cL \Lt_{-1}
+ \frac{15}{c} \cW_3 \Wt_{-2}^{(3)}
+ \frac{28}{c} \cW_4 \Wt_{-3}^{(4)}
+ \frac{45}{c} \cW_5 \Wt_{-4}^{(5)}
 , \nonumber \\
a_{t}(t,\phi)&=
  \mu_{2}  \Lt_{1}
+ \mu_{3}  \Wt_{2}^{(3)}
+ \mu_{4}  \Wt_{3}^{(4)}
+ \mu_{5}  \Wt_{4}^{(5)}\nonumber\\
+&\sum_{i=-1}^{0} \nu_{2}^{(i)}  \Lt_{i}
+ \sum_{i=-2}^{1} \nu_{3}^{(i)}  \Wt_{i}^{(3)}
+ \sum_{i=-3}^{2} \nu_{4}^{(i)}  \Wt_{i}^{(4)}
+ \sum_{i=-4}^{3} \nu_{5}^{(i)}  \Wt_{i}^{(5)}\nonumber\\
&=a_{t}^{( \mu_{2})}+a_{t}^{( \mu_{3})}+a_{t}^{( \mu_{4})}+a_{t}^{( \mu_{5})}.
\end{align}\\

Here,\,$\mu_s$'s are in principle arbitrary functions of $t$ and $\phi$.\,We can interpret these arbitrary functions  as chemical potentials.\,This means that we assume those chemical potential to be fixed at infinity,\,i.e.\,$\delta\mu_2=\delta\mu_3=\delta\mu_4=\delta\mu_5=0$.\,The remaining functions $\nu_{s}^{(i)}$'s  are fixed by the flatness condition (\ref{flatness}).\,For these fixed chemical potentials $\mu_s$'s,\,the time evolution of canonical boundary charges $\cL$ and $\cW_s$'s  as well as $\nu_{s}^{(i)}$'s is similar as in the procedure of the Sec.\,\ref{cp}.\,The final form of the temporal connection $a_t$ is given in Appendix \ref{cp5}.

In many applications,\,however,\,the chemical potentials $\mu_s$'s can be taken as a constant,\,which can simplify most of the  formulas,\,especially in $AdS_3$ higher spin gravity,\,considerably.\,Therefore,\,one note that from now on one will assume that the chemical potential $\mu_s$'s as well as the canonical boundary charge $\cL$ and $\cW_s$'s are constant.\,Under this assumption,\,the temporal connection $a_{t}$ simplify as

\begin{subequations}
  \begin{align}
a_{t}^{( \mu_{2})}&=
\mu_2 \bigg(
  \Lt_{1}
+ \frac{6}{c} \cL \Lt_{-1}
+ \frac{15}{c} \cW_3 \Wt_{-2}^{(3)}
+ \frac{28}{c} \cW_4 \Wt_{-3}^{(4)}
+ \frac{45}{c} \cW_5 \Wt_{-2}^{(5)}
\bigg)
 \end{align}
  \begin{align}
a_{t}^{( \mu_{3})}&=
\mu_3 \bigg(\Wt_2^{(3)}+\frac{12}{c} \mathcal{W}_3 \Lt_{-1} +
\bigg(\frac{36}{c^2} \mathcal{L}^2+\frac{16}{c} \sqrt{\frac{15}{7}} \mathcal{W}_4\bigg)  \Wt_{-2}^{(3)}
+\bigg(\frac{96}{c^2} \sqrt{\frac{15}{7}} \mathcal{L}\mathcal{W}_3+\frac{8 \sqrt{15}}{c} \mathcal{W}_5\bigg)  \Wt_{-3}^{(4)}\nonumber\\
&+\bigg(\frac{1800}{7 \sqrt{7} c^2} \mathcal{W}_3^2+\mathcal{W}_4 \frac{48
   \sqrt{15}}{c^2} \mathcal{L}\bigg)  \Wt_{-4}^{(5)}+\frac{8 \sqrt{15}}{c} \mathcal{W}_4  \Wt_{-2}^{(5)}
+\frac{16}{c} \sqrt{\frac{15}{7}} \mathcal{W}_3  \Wt_{-1}^{(4)}+\frac{12}{c} \mathcal{L}  \Wt_0^{(3)}\bigg)
  \end{align}
  \begin{align}
a_{t}^{( \mu_{4})}&=
\mu_4 \bigg(\Wt_3^{(4)}+\frac{18}{c} \mathcal{W}_4 \Lt_{-1} +
\bigg(\frac{108}{c^2} \mathcal{L}^2+\frac{2 \sqrt{105}}{c} \mathcal{L}^2\bigg)  \Wt_{-1}^{(4)}+\bigg(\frac{540 \sqrt{15}}{7
   c^2} \mathcal{L} \mathcal{W}_3-\frac{30}{c} \sqrt{\frac{15}{7}} \mathcal{W}_5\bigg)  \Wt_{-2}^{(5)}\nonumber\\
&+\bigg(\frac{576}{7 c^2} \sqrt{\frac{15}{7}} \mathcal{L} \mathcal{W}_3+\frac{45 \sqrt{15}}{7 c} \mathcal{W}_5\bigg)  \Wt_{-2}^{(3)}+\bigg(\frac{216}{c^3} \mathcal{L}^3+\frac{1500}{7
   c^2} \mathcal{W}_3^2+\frac{44}{c^2} \sqrt{\frac{21}{5}} \mathcal{L} \mathcal{W}_4\bigg)  \Wt_{-3}^{(4)}\nonumber\\
&+\bigg(\frac{1620 \sqrt{15}}{7 c^3} \mathcal{L}^2 \mathcal{W}_3+\frac{900}{\sqrt{7} c^2} \mathcal{W}_3 \mathcal{W}_4-\frac{18 \sqrt{105}}{c^2} \mathcal{L} \mathcal{W}_5\bigg)  \Wt_{-4}^{(5)}+\frac{45
   \sqrt{15}}{7 c} \mathcal{W}_3  \Wt_0^{(5)}+\frac{120}{7 c} \sqrt{\frac{15}{7}} \mathcal{W}_3  \Wt_0^{(3)}\nonumber\\
&+\frac{18}{c} \mathcal{L}  \Wt_1^{(4)}\bigg)
\end{align}
  \begin{align}
a_{t}^{( \mu_{5})}&=
\mu_5 \bigg( \Wt_4^{(5)}+\frac{24}{c} \mathcal{W}_5 \Lt_{-1}+
\bigg(\frac{216}{c^2} \mathcal{L}^2-\frac{4 \sqrt{105}}{c} \mathcal{W}_4\bigg)  \Wt_0^{(5)}
+\bigg(\frac{1440}{7 \sqrt{7} c^2} \mathcal{W}_3^2+\frac{240 \sqrt{15}}{7
   c^2} \mathcal{L} \mathcal{W}_4\bigg)  \Wt_{-2}^{(3)}\nonumber\\
&+\bigg(\frac{576 \sqrt{15}}{7 c^2} \mathcal{L} \mathcal{W}_3-\frac{4
   \sqrt{105}}{c} \mathcal{W}_5\bigg)  \Wt_{-1}^{(4)}+\bigg(\frac{864}{c^3} \mathcal{L}^3+\frac{2400}{7
   c^2} \mathcal{W}_3^2-\frac{272}{c^2} \sqrt{\frac{15}{7}} \mathcal{L} \mathcal{W}_4\bigg)  \Wt_{-2}^{(5)}\nonumber\\
&+\bigg(\frac{1440 \sqrt{15}}{7
   c^3} \mathcal{L}^2 \mathcal{W}_3+\frac{112
   \sqrt{7}}{c^2} \mathcal{W}_3 \mathcal{W}_4-\frac{104}{c^2} \sqrt{\frac{15}{7}} \mathcal{L} \mathcal{W}_5\bigg)  \Wt_{-3}^{(4)}+\bigg(\frac{1296}{c^4} \mathcal{L}^4
   -\frac{624}{c^3} \sqrt{\frac{15}{7}} \mathcal{L}^2 \mathcal{W}_4\nonumber\\
&+\frac{90000}{49 c^3} \mathcal{L} \mathcal{W}_3^2+\frac{420}{c^2} \mathcal{W}_4^2-\frac{990}{\sqrt{7}
   c^2} \mathcal{W}_3 \mathcal{W}_5\bigg)  \Wt_{-4}^{(5)}+\frac{8 \sqrt{15}}{c} \mathcal{W}_3  \Wt_1^{(4)}+\frac{8
   \sqrt{15}}{c} \mathcal{W}_4  \Wt_0^{(3)}+\frac{24}{c} \mathcal{L}  \Wt_2^{(5)} \bigg)
\end{align}
\end{subequations}
This manifestly solve the field equation\,(\ref{flatness}),\,and describe black holes solutions carrying not only mass and angular momentum but also canonical charges of the gravity.\,As discussed in Ref.\,\onlinecite{Henneaux:2013dra},\,and comparison with e.g.\,eqn.\,(2.29) of Ref.\,\onlinecite{Bunster:2014mua} shows that  the temporal component of the connection is linear  in the chemical potentials $\mu_s$'s.\,The same procedure can be followed to establish $\bar{a}$ with a constant chemical potential.\,If one set $ \mu_3=\mu_4=\mu_5=0$ and $\cW_s=0;\,(s=3,4,5)$,\,this connection reduces to the usual $BTZ$ black hole.\,This solution is therefore interpreted as a generalization of the $BTZ$ black hole to include nonzero spin s:\,$(s\,=\,3,\,4,\,5)$ charges and related chemical potentials $ \mu_3,\,\mu_4$ and $\mu_5$.
\subsection{Determining the $1/c$-dependence of the structure constants}\label{qc}
So far,\,the asymptotic  symmetry algebra we obtained is not the quantum algebra because of the appearance of the nonlinear terms in the $\mathcal{W}_5$ algebra,\,but rather one semi-classical algebra.\,We know that this fits with the fact that classical limit takes $ c\rightarrow \infty$.\,For this,\,we are interested in the quantum mechanical version of the asymptotic symmetry  algebra, hence we take into account normal ordering effects.\,We also emphasize here that we know that the normal ordered version of the algebra  then becomes inconsistent for finite values of the central charge $c$ and thus has to be modified.\,So we have to lift the semi-classical results to quantum results and finally end up with the quantum $\mathcal{W}_5$ asymptotic symmetry algebra.\,Finally,\,a deformed version of usual semi-classical asymptotic symmetry algebra with new arbitrary structure constants can be obtained.\, Refs.\,\onlinecite{Ozer:2002xe,Hornfeck:1992tm}\,This will be the final form of the quantum asymptotic symmetry algebra of the $AdS_3$ higher spin gravity.

In the shorthand notation,\,all important  structure constants  $\cC_{s_1 s_2}^{s_3}$'s appearing in the classical asymptotic $\mathcal{W}_5$ symmetry algebra as in Appendix \ref{appE} are given by schematically
\be
\begin{split}
 \mathcal{W}_{3}\ast \mathcal{W}_{3}\,\,&\sim \,\, {{c}\over {3}}\,\bf{1}\,+\,\cC_{33}^{4}\mathcal{W}_{4}\\
 \mathcal{W}_{3}\ast \mathcal{W}_{4}\,\,&\sim \,\, \cC_{34}^{3}\mathcal{W}_{3} \,+\,\cC_{34}^{5}\mathcal{W}_{5}\\
 \mathcal{W}_{4}\ast \mathcal{W}_{4}\,\,&\sim \,\, {{c}\over {4}}\,\bf{1} \,+\,\cC_{44}^{4}\mathcal{W}_{4}\,+\,\cC_{44}^{[33]}[\mathcal{W}_{3}\mathcal{W}_{3}]\\
 \mathcal{W}_{3}\ast \mathcal{W}_{5}\,\,&\sim \,\, \cC_{35}^{4}\mathcal{W}_{4} \,+\,\cC_{35}^{[33]}[\mathcal{W}_{3}\mathcal{W}_{3}]\\
 \mathcal{W}_{4}\ast \mathcal{W}_{5}\,\,&\sim \,\, \cC_{45}^{3}\mathcal{W}_{3} \,+\,\cC_{45}^{5}\mathcal{W}_{5} \,+\,\cC_{45}^{[34]}[\mathcal{W}_{3}\mathcal{W}_{4}] \,+\,\cC_{45}^{[34]'}\partial[\mathcal{W}_{3}\mathcal{W}_{4}]\\
 \mathcal{W}_{5}\ast \mathcal{W}_{5}\,\,&\sim \,\, {{c}\over {5}}\,\bf{1} \,
+\,\cC_{55}^{4}\mathcal{W}_{4}\,
+\,\cC_{55}^{[33]}[\mathcal{W}_{3}\mathcal{W}_{3}]
+\,\cC_{55}^{[35]}[\mathcal{W}_{3}\mathcal{W}_{5}]\\
&+\,\cC_{55}^{[44]}[\mathcal{W}_{4}\mathcal{W}_{4}]
+\,\cC_{55}^{[33]''}\partial^{2}[\mathcal{W}_{3}\mathcal{W}_{3}]\\
\end{split}
\ee
where we only write out the contributions of the Virasoro primaries and the composite primary fields to the singular part of OPEs.\,We must also emphasize here that the structure constants $\cC _{s_1 s_2}^{s_3}$'s displayed in (57) are sufficient to obtain all the remaining one,\,especially given that in the semiclassical case all structure constants have been displayed explicitly. Nevertheless,\,some important structure constants  for classical and quantum asymptotic $\mathcal{W}_5$ symmetry algebra are given by
\be
\begin{split}
&\cC_{34}^{3}\,=\,\frac{3}{4}\cC_{33}^{4},\,\,\
 \cC_{35}^{4}\,=\,\frac{4}{5}\cC_{34}^{5},\,\,\
 \cC_{45}^{3}\,=\,\frac{3}{5}\cC_{34}^{5}\,=\,\frac{3}{4}\cC_{35}^{4},\,\,\
 \cC_{55}^{4}\,=\,\frac{4}{5}\cC_{45}^{5}\\
\end{split}
\ee
\begin{center}
\begin{tabular}{|c|c|c|}
\hline
 & classical algebra & quantum algebra \\
\hline
$\cC_{33}^{4}$ & $\frac{32}{\sqrt{105}}$  &$\sqrt{\frac{1024 (c+2) (c+23)}{3 (5 c+22) (7 c+68)}}$  \\
\hline
$\cC_{34}^{5}$ & $\frac{5\sqrt{15}}{7}$  &$\sqrt{\frac{25 (3 c+116) (5 c+22)}{(7 c+68) (7 c+114)}}$  \\
\hline
$\cC_{45}^{5}$ & $\frac{11}{14}\frac{\sqrt{15}}{7}$  &$\sqrt{\frac{75 \left(11 c^3+204 c^2+9340 c+70272\right)^2}{4 (c+2) (c+23) (5 c+22) (7 c+68) (7 c+114)^2}}$  \\
\hline
$\cC_{44}^{[33]}$ &$\frac{45}{2c}$  &$\frac{9 (5 c+22)}{2 (c+2) (c+23)}$  \\
\hline
$\cC_{35}^{[33]}$ &$\frac{24}{\sqrt{7}c}$  &$\sqrt{\frac{432 (2 c-1)^2}{(c+2) (c+23) (3 c+116) (7 c+114)}}$  \\
\hline
$\cC_{45}^{[34]}$ &$\frac{74}{\sqrt{7}c}$  &$\sqrt{\frac{12 (37 c+334)^2}{(c+2) (c+23) (3 c+116) (7 c+114)}}$  \\
\hline
$\cC_{45}^{[34]'}$ &$\frac{66}{\sqrt{7}c}$  &$\sqrt{\frac{108 (3 c+116)}{(c+2) (c+23) (7 c+114)}}$  \\
\hline
$\cC_{55}^{[33]}$ &$\frac{181}{14c}$  &$\frac{3 (181 c^3+14880 c^2+248948 c+1507824)}{2 (c+2) (c+23) (3 c+116) (7 c+114)}$  \\
\hline
$\cC_{55}^{[44]}$ &$\frac{448}{15c}$  &$\frac{64 (7 c+114)}{(3 c+116) (5 c+22)}$  \\
\hline
\end{tabular}\\
\item Table-1: some important  structure constants for classical and  \\
 quantum asymptotic $\mathcal{W}_5$ symmetry algebra with $1/c$ correction.
\end{center}
One can check as in the Table-1 that if it can be  taken into account the normal ordering effects in the quantum asymptotic $\mathcal{W}_5$ symmetry algebra,\,then the classical asymptotic $\mathcal{W}_5$ symmetry algebra can be obtained under  $ c\rightarrow \infty$ classical limit with $1/c$ correction as in the second column of the Table-1,\,i.e $\cC_{33}^{4}$

\be
\begin{split}
\left(\cC_{33}^{4}\right)^2\,&=\,\frac{1024 (c+2) (c+23)}{3 (5 c+22) (7 c+68)}\\
              &=\,\frac{1024+\frac{25600}{c}+\frac{47104}{c^2} }{105+\frac{1482}{c}+\frac{4488}{c^2}}\\
\end{split}
\ee
under  $ c\rightarrow \infty$ classical limit with $1/c$ correction
\be
\begin{split}
\cC_{33}^{4}\,&=\,\frac{32}{\sqrt{105}}\,+\,\mathcal{O}\left(1\over c\right),
\end{split}
\ee
which is compatible with Table-1.

Finally,\,one can calculate the Operator Product Expansions of the quantum $\mathcal{W}_N$  asymptotic symmetry algebra by starting with the classical one of Chern-Simons theory based on $\mathfrak{sl}(N,\bR)$ Lie algebra.\,The corrections for finite $c$ in the nonlinear terms can also be determined recursively by solving some constraints for the Jacobi identities given in
the Appendix \ref{appE} and Ref.\,\onlinecite{Ozer:2002xe},\,especially for $\mathfrak{sl}(5,\bR)$ Lie algebra.

\subsection{Holonomies in  $\mathfrak{sl}(5,\bR) \oplus \mathfrak{sl}(5,\bR)$ Chern\,-\,Simons Theory }\label{hol1}
The $\mathfrak{sl}(5,\bR)$ higher spin  black hole solution will be characterised by independent global charges  $\mathcal{W}_s$ and the parameters
$\mu_s$,$(s\,=\,2,\,3,\,4,\,5)$ are the potentials conjugate to these charges.\,For $s\,=\,2$,\,$\mathcal{W}_2$ is the stress tensor $\cL$ in the boundary
and $\mu_2$ is the inverse temperature $\beta$.\,The charges are again recovered from the partition function
\begin{align}\label{pp1}
\cZ=\mathfrak{tr}\bigg( e^{4\pi^2 i\sum_{s=2}^5\mu_s\mathcal{W}_s}\bigg)
\end{align}
or taking the logarithmic of the partition function we find
\begin{align}\label{pp2}
\mathfrak{ln}\cZ=\cS+4\pi^2 i\sum_{s=2}^5\mu_s\mathcal{W}_s
\end{align}
as
\begin{align}\label{spins}
 \mathcal{W}_s=-\frac{i}{4\pi^2 } \frac{\partial\mathfrak{ln}\cZ }{\partial \mu_s},
\end{align}
or equivalently
\begin{align}\label{pots}
\mu_s=\frac{i}{4\pi^2 } \frac{\partial\cS}{\partial \mathcal{W}_s},
\end{align}
A novel feature however,\,is that these expressions are seen to satisfy:
\begin{align}\label{intc1}
\frac{\partial\mathcal{W}_s }{\partial \mu_{s'}}=\frac{\partial\mathcal{W}_{s'} }{\partial \mu_s},\,\,s\neq {s'}.
\end{align}
This is called the \textit{integrability conditions} which the black hole solution has to satisfy in order to associate to the black hole a partition function as in (\ref{pp1}),\,and similar expressions for the barred sector.\,Another way to say this is that  needs to be satisfied in order that (\ref{intc1}) the thermodynamic quantities assigned
to the black hole will obey the first law of thermodynamics.\\
As in Appendix \ref{appB} for $\mathfrak{sl}(3,\bR) \oplus \mathfrak{sl}(3,\bR)$ case,\,let's recall the connection $a_\phi$ for
$\mathfrak{sl}(5,\bR) \oplus \mathfrak{sl}(5,\bR)$ Chern\,-\,Simons Theory as:
\begin{equation}
		a_\phi=\left(
\begin{array}{ccccc}
 0 & \frac{\cQ_2}{10} & \frac{\cQ_3}{7 \sqrt{6}} & \frac{\cQ_4}{24}-\frac{17 \cQ_2^2}{1200} & \frac{\cQ_5}{24}-\frac{31 \cQ_2 \cQ_3}{840} \\
 2 & 0 & \frac{1}{10} \sqrt{\frac{3}{2}} \cQ_2 & \frac{\cQ_3}{14} & \frac{\cQ_4}{24}-\frac{17 \cQ_2^2}{1200} \\
 0 & \sqrt{6} & 0 & \frac{1}{10} \sqrt{\frac{3}{2}} \cQ_2 & \frac{\cQ_3}{7 \sqrt{6}} \\
 0 & 0 & \sqrt{6} & 0 & \frac{\cQ_2}{10} \\
 0 & 0 & 0 & 2 & 0 \\
\end{array}
\right)
\end{equation}
depending on \textit{Casimir operator}-related parameters $\cQ_s=\frac{1}{s}\mathfrak{tr}[a_\phi^s]$,\,($s\,=\,2,\,3,\,4,\,5)$ for $\mathfrak{sl}(5,\bR)$ Lie algebra.\,The easiest way to solve the holonomy condition is by solving the characteristic polynomial.\,This means that one may express exponentiated matrix $w=2\pi a_\phi$,\,a general $5\times5$ matrix as in (\ref{holmat2}).
\begin{equation}\label{holmat5}
w^5\,=\,\Theta_0+\Theta_1 w+\Theta_2 w^2+\Theta_3 w^3+\Theta_4 w^4
\end{equation}
To find the \textit{holonomy invariants} $\Theta$'s one consider the general characteristic equation of exponentiated matrix $w$,\,explicitly
\begin{equation}\label{holmat4}
\lambda ^5\,=\,4 \pi ^2 \lambda ^3 \cQ_2+8 \pi ^3 \lambda ^2 \cQ_3-8 \pi ^4 \lambda  \cQ_2^2+16 \pi ^4 \lambda  \cQ_4-32 \pi ^5 \cQ_2 \cQ_3+32 \pi ^5 \cQ_5
\end{equation}
If $\lambda_i$ is $i^{th}$ eigenvalue of the exponentiated matrix $w$,\,and identify the \textit{holonomy invariants} in (\ref{holmat5}) as
\begin{align}\label{thetas}
&\Theta_0=\mathfrak{det}[w]=32 \pi ^5 \left(Q_5-Q_2 Q_3\right)=\sum_{i<j<k<l<m}\lambda_i\lambda_j\lambda_k\lambda_l\lambda_m\\
&\Theta_1=\frac{1}{4}\mathfrak{tr}[w^4]-\frac{1}{8}\mathfrak{tr}[w^2]^2=8 \pi ^4 \left(2 Q_4 -Q_2^2 \right)=-\sum_{i<j<k<l}\lambda_i\lambda_j\lambda_k\lambda_l\\
&\Theta_2=-\frac{1}{3}\mathfrak{tr}[w^3]=8 \pi ^3 Q_3=\sum_{i<j<k}\lambda_i\lambda_j\lambda_k  \\
&\Theta_3=-\frac{1}{2}\mathfrak{tr}[w^2]=4 \pi ^2 Q_2 =-\sum_{i<j}\lambda_i\lambda_j  \\
&\Theta_4=\mathfrak{tr}[w]=0=\sum_i\lambda_i
\end{align}
If a holonomy is trivial around a contractible cycle  the exponentiated matrix $\mathrm{e}^w$ must have eigenvalues ${0,\pm 2\pi i,\pm 4\pi i}$,\,it follows that
\begin{align}\label{holonomies2}
&\mathfrak{det}[w]=0,\,\,\mathfrak{tr}[w]=0,\,\,\mathfrak{tr}[w^3]=0,\,\,\mathfrak{tr}[w^5]=0,\\
&\mathfrak{tr}[w^2]+40 \pi^2=0,\,\,\mathfrak{tr}[w^4]-544 \pi^4=0.\nonumber
\end{align}
Now when it comes to calculating the holonomy about contractible (thermal) cycle one may use the temporal  component
\begin{align}
a_{t}=\mu_2a_{\phi}+ \sum_{i=3}^{5}\mu_{i}\bigg(a_{\phi}^{i-1}-\frac{\mathbb{I}}{5}  \mathfrak{tr}[a_{\phi}^{i-1}] \bigg)
\end{align}
 which  fulfils the equation of motion $[a_t,a_\phi]=0$  it is obvious that $a_t= \mathit{f}(a_\phi)$.\,Therefore the trivial holonomy is
\begin{equation}\label{hol2}
H_{\t}(a)\,=\,\mathrm{b^{-1}}\mathrm{e}^w\mathrm{b},
\end{equation}
where $w\,=\,\oint_{\mathcal{C}}a_{t}\mathrm{d}{t}\,=\,2\pi\tau a_{t} $ is the holonomy matrix with $\tau=\frac{i\beta}{2\pi}$.\,One can also show that this holonomy matrix gives the same holonomy conditions ($\ref{holonomies2}$) for the time circle with $w\,=\,2\pi\tau a_{t} $ and similar expressions for the barred sector.\,Thus our final goal might be to  solve the holonomy conditions  to see whether this gives a consistent  thermodynamics of the black hole with the phase structure.\,Therefore,\,the six equations of the  holonomy conditions (\ref{holonomies2}) can be explicitly written in terms of the eight parameters of the connection (\ref{spins}-\ref{pots}).\,To do this,\,the entropy function can be written in the form
\begin{align}\label{entropy1}
\cS(\cQ_2,\cQ_3,\cQ_4,\cQ_5)\,=\,2\pi k\sqrt{\cQ_2}\,\digamma(x,y,z)
\end{align}
with\,\,$\digamma(0,0,0)\,=1\,$,\,such that
\begin{align}\label{entropy2}
\mu_s\,=\,&i\frac{1}{2\pi k}\,\frac{\partial\cS}{\partial \cQ_s},\,\,(s\,=\,2,\,3,\,4,\,5)
\end{align}
where we  define three dimensionless parameters $x\,=\,\frac{\cQ_3}{\cQ_2^{3/2}}$,\,\,$y\,=\,\frac{\cQ_4}{\cQ_2^2}$  and  $z\,=\,\frac{\cQ_5}{\cQ_2 \cQ_3}$.\,Then the simplest  holonomy equation becomes
\begin{align}\label{pp1}
1 =& \digamma ^2-\left(3 x \digamma_x+4 y \digamma_y+2 z \digamma_z\right)^2\nonumber \\
   &+\frac{8}{5} (5 y-1) \digamma _x^2 +\frac{4}{5} x (25 z-6) \digamma _x \digamma _y+\frac{2}{5} \left(6 x^2+30 y-5\right) \digamma _y^2\nonumber \\
   &+\frac{2}{5} \left(46 y-50 z^2+82 z-35\right) \digamma _y \digamma _z+\frac{4}{5 x} \left(15 x^2+y (22-20 z)+4 z-5\right) \digamma _x \digamma _z\nonumber\\
   &+\frac{2}{5 x^2} \left(10 \left(x^2+1\right) z+4 (5 y-1) z^2-44 y z+4 y (y+5)-5\right)\digamma _z^2.
\end{align}
But it might be cumbersome or even impossible to solve these equations for now,\,since they're non-linear partial differential equations.\,Therefore,\,we are thinking that this calculation can make up the content of one future article,\,and now we do not go into detail.\,Finally,\,in this context,\,one can check the entropy for the $\mathfrak{sl}(5,\bR) \oplus \mathfrak{sl}(5,\bR)$ Lie algebra valued Chern-Simons gauge theory,\,which should be given in full generally by the total boundary term,\,as
\begin{equation}\label{entropy1}
S\,=\,-2\pi i k\,\mathfrak{tr}(a_{t}a_\phi)\,=\,-2\pi i k\,\sum_{n=2}^5 n \mu_n \cQ_n
\end{equation}
and similar expressions for the barred sector.
\section{Summary and Conclusion}\label{sum}

In this work,\,we first reviewed a relation between $AdS_3$ and  $\mathfrak{sl}(2,\bR) \oplus \mathfrak{sl}(2,\bR)$ Chern\,-\,Simons theory.\,The Chern\,-\,Simons formulation of $AdS_3$ allows for a straightforward generalization to a higher spin theory.\,The higher spin gauge fields have no propagating degree of freedom,\,but we noted that there are  a large class of interesting non-trivial solution.\,Specifically,\,$AdS_3$ in the presence of a tower of higher spin fields up to spin 5 is obtained by enlarging $\mathfrak{sl}(2,\bR) \oplus \mathfrak{sl}(2,\bR)$ to $\mathfrak{sl}(5,\bR) \oplus \mathfrak{sl}(5,\bR)$.\,Finally,\,classical $\mathcal{W}_5$ symmetry algebra as spin 5 asymptotic symmetry algebra is obtained.

As in the Sec.\,\ref{hscs}, the asymptotic $\mathcal{W}_5$ symmetry algebra we obtained is not the final form of the related algebra of the $AdS_3$ spin 5 gravity since it is just effectively for the Virasoro central charge $c$ in the large values.\,On the other hand,\,we also indicated how to introduce chemical potentials and holonomy conditions associated with these higher spin charges in higher spin $AdS_3$ gravity in a manner that it preserves the asymptotic symmetry algebra. \cite{deBoer} However,\,from $CFT_2$ point of view,\,it is not clear how the introductions of the chemical potentials  through the time component can be interpreted.\,One can proceed entirely in the same way for $AdS_3$ spin 5 gravity and thus\,determine the asymptotic symmetry algebra as quantum $\mathcal{W}_5$ algebra instead of  rather than semiclassical one.\,Therefore,\,we can take into account normal ordering effects of the quantum $\mathcal{W}_5$  algebra as in Ref.\,\onlinecite{Ozer:2002xe}.\,So far we only discussed the classical $\mathcal{W}_5$ symmetry.\,One may wonder,\,whether the quantum case also admits a description in terms of an classical $\mathcal{W}_N$ symmetry,\,or a quantum version of it.\,These questions seem to fairly non-trivial,\,but it seems that the procedure does go through  straightforwardly.
\section{Acknowledgments}
We thank Marc Henneaux for both his encouragement and constructive discussions on the manuscript.\,This work was supparted by Istanbul Technical University Scientific Research Projects Department(ITU BAP,\,project number:40199).

\appendix
\section{$\mathfrak{sl}(2,\bR)$ generators}\label{appA}
Below,\,we denote the three generators of $\mathfrak{sl}(2,\bR)$ Lie algebra,\,which  will be assumed to be described by a set of matrices $\tt L_{a}$ with $\,a\,=\,\pm1,0$\, given by
\begin{alignat}{1}
\tt  L_0\,&=\,\frac{1}{2}
\begin{pmatrix}
 1 & 0 \\
 0 &-1
\end{pmatrix} ,
\tt  L_1 \, = \,
\begin{pmatrix}
 0 & 0\\
  -1 & 0
\end{pmatrix} ,
\tt  L_{-1} \, = \,
\begin{pmatrix}
 0 & 1\\
  0 & 0
\end{pmatrix} ,
\end{alignat}
which admit an invariant bilinear form
\begin{equation}
		\eta_{ab}=\left(
			\begin{array}{c|ccc}
				  &\tt L_{-1}&\tt L_0&\tt L_{1}\\
				\hline
				\tt L_{-1}&0&0&-1\\
				\tt L_0&0&\frac{1}{2}&0\\
				\tt L_{1}&-1&0&0
			\end{array}\right),
	\end{equation}

\section{$\mathfrak{sl}(3,\bR) \oplus \mathfrak{sl}(3,\bR)$  Higher spin  Chern\,-\,Simons theory}\label{appB}
The spin-3 case is given by promoting $\mathfrak{sl}(2,\bR)$ to $\mathfrak{sl}(3,\bR)$.\,
\subsection{$\mathfrak{sl}(3,\bR)$ generators as the principal embedding of $\mathfrak{sl}(2,\bR)$ }\label{app32}
To construct this principal embedding,\,we denote the eight generators of $\mathfrak{sl}(3,\bR)$ Lie algebra,\,which will be assumed to be described by a set of matrices.\,The principal embedding of $\mathfrak{sl}(2,\bR)$ Lie algebra to $\mathfrak{sl}(3,\bR)$ Lie algebra contains the spin 2 triplet $\Lt_{n}$ with $n=\pm1,0$, and the spin 3 triplet $\Wt_{n}^{(3)}$  and can be represented in terms of the  commutation relations for $\Lt_{n}\equiv\Wt_{n}^{(2)}$.\,Below,\,we collect the eight $\mathfrak{sl}(3,\bR)$ generators:

\begin{alignat}{1}
\tt  L_0\,&=\,\frac{1}{2}
\begin{pmatrix}
  1 & 0 & 0 \\
 0 & 0 & 0 \\
 0 & 0 & -1 \\
\end{pmatrix} ,
\tt  L_1 \, = \,\sqrt{2}
\begin{pmatrix}
 0 & 0 & 0 \\
 1 & 0 & 0 \\
 0 & 1 & 0 \\
\end{pmatrix} ,
\tt  L_{-1} \, = \,\sqrt{2}
\begin{pmatrix}
 0 & -1 & 0 \\
 0 & 0 & -1 \\
 0 & 0 & 0 \\
\end{pmatrix} ,  \nn \\
\tt W_{0}^{(3)}&=2\sqrt{-\frac{\sigma}{3} }
\begin{pmatrix}
 1 & 0 & 0 \\
 0 & -2 & 0 \\
 0 & 0 & 1 \\
\end{pmatrix} ,
\tt W_{1}^{(3)}\, =\, \sqrt{-6 \sigma }
\begin{pmatrix}
 0 & 0 & 0 \\
 1 & 0 & 0 \\
 0 & -1 & 0 \\
\end{pmatrix} ,
\tt W_{2}^{(3)}\, =\, 4\sqrt{-3 \sigma }
\begin{pmatrix}
 0 & 0 & 0 \\
 0 & 0 & 0 \\
 6 & 0 & 0 \\
\end{pmatrix} , \nn\\
\end{alignat}
 and $\tt W_{-m}^{(s)}\,=\,(-1)^m {\tt W_{m}^{(s)}}^\dag$

\subsection{Holonomies in  $\mathfrak{sl}(3,\bR) \oplus \mathfrak{sl}(3,\bR)$ Chern\,-\,Simons Theory }\label{hol33}
Let's recall the connection $a_\phi$ for  $\mathfrak{sl}(3,\bR) \oplus \mathfrak{sl}(3,\bR)$ Chern\,-\,Simons Theory as:
  \begin{eqnarray}
a_{\phi}\,=\,\left(
\begin{array}{ccc}
 0 & \frac{\cQ_2}{2 \sqrt{2}} & \frac{\cQ_3}{2} \\
 \sqrt{2} & 0 & \frac{\cQ_2}{2 \sqrt{2}} \\
 0 & \sqrt{2} & 0 \\
\end{array}
\right)
\end{eqnarray}
depending on \textit{Casimir operator}-related parameters $\cQ_s=\frac{1}{s}\mathfrak{tr}[a_\phi^s]$,\,($s\,=\,2,\,3)$ for $\mathfrak{sl}(3,\bR)$ Lie algebra.\,Now when it comes to calculating the holonomy about contractible (thermal) cycle one may be used the temporal  component as:
\begin{align}
a_{t}=\mu_2a_{\phi}+ \mu_{3}\bigg(a_{\phi}^{2}-\frac{\mathbb{I}}{3}  \mathfrak{tr}[a_{\phi}^{2}] \bigg)
\end{align}
which  fulfils the equation of motion $[a_t,a_\phi]=0$  it is obvious that $a_t= \mathit{f}(a_\phi)$.\,Therefore the trivial holonomy is
\begin{equation}\label{hol2}
H_{\t}(a)\,=\,\mathrm{b^{-1}}\mathrm{e}^w\mathrm{b},
\end{equation}
where $w\,=\,\oint_{\mathcal{C}}a_{t}\mathrm{d}{t}\,=\,2\pi\tau a_{t} $ is the holonomy matrix with $\tau=\frac{i\beta}{2\pi}$.\,If a holonomy is trivial around a contractible cycle  the exponentiated matrix $\mathrm{e}^w$ must have eigenvalues ${0,\pm 2\pi i}$,\,it follows that
\begin{align}\label{holonomies2}
\mathfrak{det}[w]=0,\,\,\mathfrak{tr}[w^2]+8 \pi^2=0.
\end{align}
These conditions are written as:
\begin{eqnarray}
&0\,=&\,27 \mu _2^3 \cQ_3+18 \mu _3 \mu _2^2 \cQ_2^2+27 \mu _3^2 \mu _2 \cQ_2 \cQ_3-2 \mu _3^3 \cQ_2^3+27 \mu _3^3 \cQ_3^2\\
&0\,=&\,3+3 \mu _2^2 \cQ_2 \tau ^2+\mu _3^2 \cQ_2^2 \tau ^2+9 \mu _2 \mu _3 \cQ_3 \tau ^2.\label{holcon2}
\end{eqnarray}
In order to  solve the holonomy conditions,\,the entropy function can be written in the form
\begin{align}\label{entropy1}
\cS(\cQ_2,\cQ_3)\,=\,2\pi k \sqrt{\cQ_2}\, \digamma(z)
\end{align}
with\,\,$\digamma(0)\,=\,1$,\,we define a dimensionless parameter $z\,=\,\frac{27}{2}\frac{\cQ_3}{\cQ_2^{3/2}}$ such that
\begin{align}\label{entropy2}
\mu_s\,=\,&i\frac{1}{2\pi k }\, \frac{\partial\cS}{\partial \cQ_s},\,\,(s\,=\,2,\,3).
\end{align}
Then the  holonomy equations become
\begin{align}\label{pp1}
1\,=\,& \digamma^2-36 z (z-2)(\digamma^{'})^2\nonumber,\\
0\,=\,& \digamma ^3- 18 (z-2) \digamma ' \left(6 z \digamma ' \left(2 (z-2) \digamma '-\digamma \right)+\digamma ^2\right)
\end{align}
and similar expressions for the barred sector.\,This is compatible with the seminal work of Gutperle and Kraus. \cite{Gutperle:2011kf}
\section{$\mathfrak{sl}(5,\bR)$ generators as the principal embedding of $\mathfrak{sl}(2,\bR)$ }\label{appC}
Throughout this work we used the following matrix representations of $\mathfrak{sl}(5,\bR)$ Lie algebra.\,Below,\,we denote the twenty-four generators of $\mathfrak{sl}(5,\bR)$ Lie algebra,\,which  will be assumed to be described by a set of matrices in Sec.\,\ref{sec:first-commutators}.\,The principal embedding of $\mathfrak{sl}(2,\bR)$ Lie algebra to $\mathfrak{sl}(5,\bR)$ Lie algebra contains the spin 2 triplet $\Lt_{n}$ with
$n=\pm1,0$,\,the spin 3 triplet $\Wt_{n}^{(3)}$ with $n=\pm2,\pm1,0$,\,the spin 4 quartet $\Wt_{n}^{(4)}$ with $n=\pm3,\pm2,\pm1,0$ and the spin 5 quintet $\Wt_{n}^{(5)}$ with $n=\pm4,\pm3,\pm2,\pm1,0$ and can be represented in terms of the  commutation relations for $\Lt_{n}\equiv\Wt_{n}^{(2)}$ as follows

\begin{subequations}
  \begin{align}
    [\Lt_{n},\Wt_{m}^{(s)}]&=((s-1)n-m) \Wt^{(s)}_{n+m}, (s=2,3,4,5)
      \\
    [\Wt_{n}^{(3)},\Wt_{m}^{(3)}]&=f_{33}^{2}(n,m)\Lt_{n+m}+ f_{33}^{4}(n,m) \Wt_{n+m}^{(4)}
      \\
    [\Wt_{n}^{(3)},\Wt_{m}^{(4)}]&=f_{34}^{3}(n,m)\Wt_{n+m}^{(3)}+ f_{34}^{5}(n,m) \Wt_{n+m}^{(5)}
      \\
    [\Wt_{n}^{(4)},\Wt_{m}^{(4)}]&=f_{44}^{2}(n,m)\Lt_{n+m}+ f_{44}^{4}(n,m) \Wt_{n+m}^{(4)}
      \\
    [\Wt_{n}^{(3)},\Wt_{m}^{(5)}]&=f_{35}^{4}(n,m) \Wt_{n+m}^{(4)}
     \\
    [\Wt_{n}^{(4)},\Wt_{m}^{(5)}]&=f_{45}^{3}(n,m)\Wt_{n+m}^{(3)}+ f_{45}^{5}(n,m) \Wt_{n+m}^{(5)}
      \\
    [\Wt_{n}^{(5)},\Wt_{m}^{(5)}]&=f_{55}^{2}(n,m)\Lt_{n+m}+ f_{55}^{4}(n,m) \Wt_{n+m}^{(4)}.
  \end{align}
\end{subequations}
where
\begin{subequations}
  \begin{align}
 f_{33}^{2}(n,m)\,=\,&\sigma_1\,(n-m)(2 n^2 - n m + 2 m^2 - 8)\\
 f_{33}^{4}(n,m)\,=\,&\sigma_2\,(n-m)\\
 f_{34}^{3}(n,m)\,=\,&\sigma_3\,(m^3 -3 m^2 n+ 5 m n^2 -9 m -5 n^3+ 17 n)\\
 f_{34}^{5}(n,m)\,=\,&\sigma_4\,(3 n-2 m)\\
 f_{44}^{2}(n,m)\,=\,&\sigma_5\,(n-m)(3 m^4 -2 m^3 n+ 4 m^2 n^2 -39 m^2 -2 m n^3+ 20 m n+ 3 n^4 -39 n^2+ 108)\\
 f_{44}^{4}(n,m)\,=\,&\sigma_6\,(n-m)(m^2 + n^2 - m n-7 )\\
 f_{35}^{4}(n,m)\,=\,&\sigma_7\,(2 m^3 -9 m^2 n+ 21 m n^2 -32 m -28 n^3+ 88 n)\\
 f_{45}^{3}(n,m)\,=\,&\sigma_8\,(3 m^5 -10 m^4 n+ 20 m^3 n^2 -75 m^3 -30 m^2 n^3+ 220 m^2 n+ 35 m n^4\nonumber\\
&-355 m n^2+ 432 m -28 n^5+ 340 n^3 -792 n)\\
 f_{45}^{5}(n,m)\,=\,&\sigma_9\,(5 m^3 -15 m^2 n+ 21 m n^2 -59 m -14 n^3+ 86 n)\\
 f_{55}^{2}(n,m)\,=\,&\sigma_{10}\,(n-m)(4 m^6 -3 m^5 n+ 6 m^4 n^2 -116 m^4 -4 m^3 n^3+ 79 m^3 n+  6 m^2 n^4-156 m^2 n^2\nonumber\\
&+ 976 m^2 -3 m n^5+ 79 m n^3 -508 m n+  4 n^6 -116 n^4+ 976 n^2 -2304)\\
 f_{55}^{4}(n,m)\,=\,&\sigma_{11}\,(n-m)(14 m^4 -21 m^3 n+ 29 m^2 n^2+ -310 m^2 -21 m n^3+ 315 m n+ 14 n^4\nonumber\\
&-310 n^2+ 1376)
 \end{align}
\end{subequations}
with
\begin{subequations}
  \begin{align}
& \sigma _3\,=\,-\frac{64 \sigma _1}{49 \sigma _2} ,
  \sigma _5\,=\,\frac{64 \sigma _1^2}{49 \sigma _2^2},
  \sigma _6\,=\,\frac{8 \sigma _1}{7 \sigma _2} ,
  \sigma _7\,=\, -\frac{5 \sigma _1}{49 \sigma _4},
  \sigma _8\,=\, -\frac{320 \sigma _1^2}{2401 \sigma _2^2 \sigma _4}\nonumber\\
& \sigma _9\,=\,\frac{8 \sigma _1}{49 \sigma _2},
  \sigma _{10}\,=\,\frac{320 \sigma _1^3}{2401 \sigma _2^2 \sigma _4^2} ,
  \sigma _{11}\,=\, -\frac{40 \sigma _1^2}{2401 \sigma _2 \sigma _4^2}
 \end{align}
\end{subequations}
where $\sigma_{i}$'s are the 11 arbitrary parameters,\,depending on three arbitrary parameters,\,mainly $\sigma_1,\sigma_2$ and $\sigma_4$ which can be changed by rescaling $\Wt_{n}^{(s)}$.\,Below,\,we collect the twenty-four $\mathfrak{sl}(5,\bR)$ generators:
\begin{alignat}{1}
\tt  L_0\,&=\,\frac{1}{2}
\begin{pmatrix}
 2 & 0 & 0 & 0 & 0 \\
 0 & 1 & 0 & 0 & 0 \\
 0 & 0 & 0 & 0 & 0 \\
 0 & 0 & 0 & -1 & 0 \\
 0 & 0 & 0 & 0 & -2 \\
\end{pmatrix} ,
\tt  L_1 \, = \,
\begin{pmatrix}
 0 & 0 & 0 & 0 & 0 \\
 2 & 0 & 0 & 0 & 0 \\
 0 & \sqrt{6} & 0 & 0 & 0 \\
 0 & 0 & \sqrt{6} & 0 & 0 \\
 0 & 0 & 0 & 2 & 0 \\
\end{pmatrix} ,
\tt  L_{-1} \, = \,
\begin{pmatrix}
 0 & -2 & 0 & 0 & 0 \\
 0 & 0 & -\sqrt{6} & 0 & 0 \\
 0 & 0 & 0 & -\sqrt{6} & 0 \\
 0 & 0 & 0 & 0 & -2 \\
 0 & 0 & 0 & 0 & 0 \\
\end{pmatrix} ,  \nn \\
\tt W_{0}^{(3)}&=\gamma_1
\begin{pmatrix}
 -2 & 0 & 0 & 0 & 0 \\
 0 & 1 & 0 & 0 & 0 \\
 0 & 0 & 2 & 0 & 0 \\
 0 & 0 & 0 & 1 & 0 \\
 0 & 0 & 0 & 0 & -2 \\
\end{pmatrix} ,
\tt W_{1}^{(3)} =\frac{\gamma_1}{2}
\begin{pmatrix}
 0 & 0 & 0 & 0 & 0 \\
 -6 & 0 & 0 & 0 & 0 \\
 0 & -\sqrt{6} & 0 & 0 & 0 \\
 0 & 0 & \sqrt{6} & 0 & 0 \\
 0 & 0 & 0 & 6 & 0 \\
\end{pmatrix} ,
\tt W_{2}^{(3)} =2\gamma_1
\begin{pmatrix}
 0 & 0 & 0 & 0 & 0 \\
 0 & 0 & 0 & 0 & 0 \\
 -\sqrt{6} & 0 & 0 & 0 & 0 \\
 0 & -3 & 0 & 0 & 0 \\
 0 & 0 & -\sqrt{6} & 0 & 0 \\
\end{pmatrix} , \nn\\
\tt W_{0}^{(4)}& =\gamma_2
\begin{pmatrix}
 -1 & 0 & 0 & 0 & 0 \\
 0 & 2 & 0 & 0 & 0 \\
 0 & 0 & 0 & 0 & 0 \\
 0 & 0 & 0 & -2 & 0 \\
 0 & 0 & 0 & 0 & 1 \\
\end{pmatrix} ,
\tt W_{1}^{(4)}=\frac{2\gamma_2}{3}
\begin{pmatrix}
0 & 0 & 0 & 0 & 0 \\
 -3 & 0 & 0 & 0 & 0 \\
 0 & \sqrt{6} & 0 & 0 & 0 \\
 0 & 0 & \sqrt{6} & 0 & 0 \\
 0 & 0 & 0 & -3 & 0 \\
\end{pmatrix} ,
\tt W_{2}^{(4)}=\frac{5\gamma_2}{3}
\begin{pmatrix}
 0 & 0 & 0 & 0 & 0 \\
 0 & 0 & 0 & 0 & 0 \\
 -\sqrt{6} & 0 & 0 & 0 & 0 \\
 0 & 0 & 0 & 0 & 0 \\
 0 & 0 & \sqrt{6} & 0 & 0 \\
\end{pmatrix} , \nn\\
%
 \tt W_{3}^{(4)}&=10\gamma_2
\begin{pmatrix}
 0 & 0 & 0 & 0 & 0 \\
 0 & 0 & 0 & 0 & 0 \\
 0 & 0 & 0 & 0 & 0 \\
 -1 & 0 & 0 & 0 & 0 \\
 0 & -1 & 0 & 0 & 0 \\
\end{pmatrix} ,
 \tt W_{0}^{(5)}=\gamma_3
\begin{pmatrix}
 1 & 0 & 0 & 0 & 0 \\
 0 & -4 & 0 & 0 & 0 \\
 0 & 0 & 6 & 0 & 0 \\
 0 & 0 & 0 & -4 & 0 \\
 0 & 0 & 0 & 0 & 1 \\
\end{pmatrix} ,
\tt W_{1}^{(5)}=\frac{5\gamma_3}{2}
\begin{pmatrix}
  0 & 0 & 0 & 0 & 0 \\
 1 & 0 & 0 & 0 & 0 \\
 0 & -\sqrt{6} & 0 & 0 & 0 \\
 0 & 0 & \sqrt{6} & 0 & 0 \\
 0 & 0 & 0 & -1 & 0 \\
\end{pmatrix}, \nn\\
%
\tt W_{2}^{(5)}&=\frac{5\gamma_3}{2}
\begin{pmatrix}
  0 & 0 & 0 & 0 & 0 \\
 0 & 0 & 0 & 0 & 0 \\
 \sqrt{6} & 0 & 0 & 0 & 0 \\
 0 & -4 & 0 & 0 & 0 \\
 0 & 0 & \sqrt{6} & 0 & 0 \\
\end{pmatrix} ,
\tt W_{3}^{(5)}=\frac{35\gamma_3}{2}
\begin{pmatrix}
  0 & 0 & 0 & 0 & 0 \\
 0 & 0 & 0 & 0 & 0 \\
 0 & 0 & 0 & 0 & 0 \\
 1 & 0 & 0 & 0 & 0 \\
 0 & -1 & 0 & 0 & 0 \\
\end{pmatrix} ,
\tt W_{4}^{(5)}=70\gamma_{3}
\begin{pmatrix}
 0 & 0 & 0 & 0 & 0 \\
 0 & 0 & 0 & 0 & 0 \\
 0 & 0 & 0 & 0 & 0 \\
 0 & 0 & 0 & 0 & 0 \\
 1 & 0 & 0 & 0 & 0 \\
\end{pmatrix},
\end{alignat}
where $\gamma_1\,=\,\sqrt{-\frac{20}{7}\sigma _1}$,\,$\gamma_2=\frac{480 \sigma _1}{7 \sigma _2}$ ,\,$\gamma_3\,=\,\frac{192}{49\sigma _2\sigma _4}\sqrt{-\frac{5}{7}\sigma _1^3}$ and $\tt W_{-m}^{(s)}\,=\,(-1)^m {\tt W_{m}^{(s)}}^\dag$,\,which admit an invariant bilinear form $\eta_{ab}$ corresponding to the values of $\sigma _i$'s specified at Sec.\,\ref{spin5}:

\begin{equation}
\tt		\eta_{ab}\,=\,\left(
\begin{array}{c|c|c|c}
\Delta_{3\times 3} & 0 &0 &0\\
  \hline
 0 & \Delta_{5\times 5} &0 &0\\
  \hline
 0 & 0 &\Delta_{7\times 7} &0\\
  \hline
 0 & 0 &0 &\Delta_{9\times 9}\\
\end{array}
\right)
\end{equation}
where
\begin{alignat}{1}
\tt\Delta_{3\times 3}\,=\,\left(
\begin{array}{ccc}
0 & 0 &-20\\
0 & 10 &0\\
-20 & 0& 0
\end{array}
\right) ,\,\,
\tt \Delta_{5\times 5}\,=\,\left(
\begin{array}{ccccc}
 0 & 0 & 0 & 0 & -8  \\
 0 & 0 & 0 & 2 & 0 \\
 0 & 0 & -\frac{4}{3} & 0 & 0  \\
 0 & 2 & 0 & 0 & 0  \\
 -8 & 0 & 0 & 0 & 0
\end{array}
\right)
        \nn \\
\end{alignat}

\newpage

\begin{alignat}{2}
\tt\Delta_{7\times 7}\,=\,\left(
\begin{array}{ccccccc}
 0 & 0 & 0 & 0 & 0 & 0 & -\frac{30}{7} \\
 0 & 0 & 0 & 0 & 0 & \frac{5}{7} & 0  \\
 0 & 0 & 0 & 0 & -\frac{2}{7} & 0 & 0  \\
 0 & 0 & 0 & \frac{3}{14} & 0 & 0 & 0  \\
 0 & 0 & -\frac{2}{7} & 0 & 0 & 0 & 0  \\
 0 & \frac{5}{7} & 0 & 0 & 0 & 0 & 0  \\
 -\frac{30}{7} & 0 & 0 & 0 & 0 & 0 & 0
\end{array}
\right) ,\,\,
\Delta_{9\times 9}\,=\,\left(
\begin{array}{ccccccccc}
 0 & 0 & 0 & 0 & 0 & 0 & 0 & 0 & -\frac{8}{3} \\
 0 & 0 & 0 & 0 & 0 & 0 & 0 & \frac{1}{3} & 0 \\
 0 & 0 & 0 & 0 & 0 & 0 & -\frac{2}{21} & 0 & 0 \\
 0 & 0 & 0 & 0 & 0 & \frac{1}{21} & 0 & 0 & 0 \\
 0 & 0 & 0 & 0 & -\frac{4}{105} & 0 & 0 & 0 & 0 \\
 0 & 0 & 0 & \frac{1}{21} & 0 & 0 & 0 & 0 & 0 \\
 0 & 0 & -\frac{2}{21} & 0 & 0 & 0 & 0 & 0 & 0 \\
 0 & \frac{1}{3} & 0 & 0 & 0 & 0 & 0 & 0 & 0 \\
 -\frac{8}{3} & 0 & 0 & 0 & 0 & 0 & 0 & 0 & 0
\end{array}
\right)
\end{alignat}

\section{Spin 5 conformal  Ward identities}\label{appD}
\subsection{Solved parameters}\label{par}
In this Appendix,\,we give a collection of eighteen solved remaining parameters as depending on $ \epsilon,\chi, \mathit{f}$ and $\eta$ parameters  which were discussed in Sec.\,\ref{spin5} of the main text.\,One finds
\begin{subequations}
\begin{align}
\chi_1&\,=\, -\chi '\\
\chi_0&\,=\,\frac{3840 \beta  \mathit{f} \sigma _ 1 \mathcal{W}_ 3}{49 \sigma _ 2}+\frac{\chi ''}{2}+\frac{460800 \gamma  \eta  \sigma _ 1^2 \mathcal{W}_ 4}{343 \sigma _ 2^2 \sigma _ 4}+2 \alpha  \chi  \mathcal{L}\\
\chi_{-1}&\,=\, -\frac{2560 \beta  \sigma _ 1 \mathcal{W}_ 3 \mathit{f}'}{49 \sigma _ 2}-\frac{1280 \beta  \mathit{f} \sigma _ 1 \mathcal{W}_ 3'}{49 \sigma _ 2}-\frac{\chi ^{(3)}}{6}-\frac{38400 \gamma  \sigma _ 1^2\nonumber \mathcal{W}_ 4 \eta '}{49 \sigma _ 2^2 \sigma _ 4}-\frac{153600 \gamma  \eta  \sigma _ 1^2 \mathcal{W}_ 4'}{343 \sigma _ 2^2 \sigma _ 4}\nonumber\\
&-\frac{2}{3} \alpha  \chi  \mathcal{L}'-\frac{5}{3} \alpha  \mathcal{L} \chi '\\
\chi_{-2}&\,=\,
\frac{832 \beta  \sigma _1 \mathcal{W}_3 \mathit{f}''}{49 \sigma _2}+\frac{960 \beta  \sigma _1 \mathit{f}' \mathcal{W}_3'}{49 \sigma _2}+\frac{3072 \alpha  \beta  \mathit{f} \sigma _1 \mathcal{W}_3 \mathcal{W}_2}{49 \sigma_2}+\frac{320 \beta  \mathit{f} \sigma _1 \mathcal{W}_3''}{49 \sigma _2}\nonumber\\
&+\frac{230400 \delta  \mathit{f} \sigma _1^2 \mathcal{W}_5}{343 \sigma _2^2 \sigma _4}
+\frac{\chi ^{(4)}}{24}+\alpha ^2 \chi\mathcal{W}_2^2+\frac{2304000 \alpha  \gamma  \eta  \sigma _1^2 \mathcal{W}_4 \mathcal{W}_2}{2401 \sigma _2^2 \sigma _4}+\frac{2}{3} \alpha  \mathcal{W}_2 \chi ''+\frac{7}{12} \alpha  \chi ' \mathcal{W}_2'\nonumber\\
&+\frac{1}{6}\alpha  \chi  \mathcal{W}_2''+\frac{92160 \beta ^2 \eta  \sigma _1^2 \mathcal{W}_3^2}{343 \sigma _2 \sigma_4}
+\beta  \mathcal{W}_3 \epsilon +\frac{556800 \gamma  \sigma _1^2 \mathcal{W}_4 \eta ''}{2401 \sigma _2^2\sigma _4}+\frac{105600 \gamma  \sigma _1^2 \eta ' \mathcal{W}_4'}{343 \sigma _2^2 \sigma _4}\nonumber\\
&+\frac{38400 \gamma  \eta  \sigma _1^2 \mathcal{W}_4''}{343 \sigma _2^2 \sigma _4}+\frac{1920 \gamma  \sigma _1 \chi\mathcal{W}_4}{49 \sigma _2}\\
\mathit{f}_2&\,=\, -\mathit{f}'\\
\mathit{f}_1&\,=\, \frac{\mathit{f}''}{2}+3 \alpha  \mathit{f} \mathcal{L}+\frac{240 \beta  \eta  \sigma _ 1 \mathcal{W}_ 3}{7 \sigma _ 4}\\
\mathit{f}_0&\,=\, -\frac{\mathit{f}^{(3)}}{6}-\frac{8}{3} \alpha  \mathcal{L} \mathit{f}'-\alpha  \mathit{f} \mathcal{L}'-\frac{180 \beta  \sigma _ 1 \mathcal{W}_ 3 \eta '}{7 \sigma _ 4}-\frac{80 \beta  \eta  \sigma _ 1 \mathcal{W}_3'}{7 \sigma _ 4}\\
\mathit{f}_{-1}&\,=\,\frac{\mathit{f}^{(4)}}{24}+\frac{7}{6}\alpha\mathcal{L}\mathit{f}''+\frac{11}{12}\alpha\mathit{f}'\mathcal{L}'+\frac{240\gamma\mathit{f}\sigma_1\mathcal{W}_4}{7\sigma_2}+3\alpha^2\mathit{f}\mathcal{L}^2+\frac{1}{4}\alpha\mathit{f}\mathcal{L}''+\frac{465\beta\sigma_1\mathcal{W}_3\eta''}{49\sigma_4}\nonumber\\
&+\frac{65\beta\sigma_1\eta'\mathcal{W}_3'}{7\sigma_4}+\frac{20\beta\eta\sigma_1\mathcal{W}_3''}{7\sigma_4}+\beta\sigma_2\chi\mathcal{W}_3-\frac{19200\delta\eta\sigma_1^2\mathcal{W}_5}{49\sigma_2\sigma_4^2}+\frac{2880\alpha\beta\eta\sigma_1\mathcal{W}_3\mathcal{L}}{49\sigma_4}\\
\mathit{f}_{-2}&\,=\,-\frac{\mathit{f}^{(5)}}{120}-\frac{1}{3}\alpha\mathit{f}^{(3)}\mathcal{L}-\frac{5}{12}\alpha\mathit{f}''\mathcal{L}'-\frac{144\gamma\sigma_1\mathcal{W}_4\mathit{f}'}{7\sigma_2}-\frac{11}{5}\alpha^2\mathcal{L}^2\mathit{f}'-\frac{7}{30}\alpha\mathit{f}'\mathcal{L}''-\frac{48\gamma\mathit{f}\sigma_1\mathcal{W}_4'}{7\sigma_2}\nonumber
\end{align}
\begin{align}
&-\frac{1}{20}\alpha\mathit{f}\mathcal{L}^{(3)}-\frac{9}{5}\alpha^2\mathit{f}\mathcal{L}\mathcal{L}'-\frac{113\beta\eta^{(3)}\sigma_1\mathcal{W}_3}{49\sigma_4}-\frac{184\beta\sigma_1\eta''\mathcal{W}_3'}{49\sigma_4}-\frac{17\beta\sigma_1\eta'\mathcal{W}_3''}{7\sigma_4}-\frac{4\beta\eta\sigma_1\mathcal{W}_3{}^{(3)}}{7\sigma_4}\nonumber\\
&-\frac{4}{5}\beta\sigma_2\mathcal{W}_3\chi'-\frac{1}{5}\beta\sigma_2\chi\mathcal{W}_3'+\frac{9600\delta\sigma_1^2\mathcal{W}_5\eta'}{49\sigma_2\sigma_4^2}+\frac{3840\delta\eta\sigma_1^2\mathcal{W}_5'}{49\sigma_2\sigma_4^2}-\frac{736\alpha\beta\eta\sigma_1\mathcal{W}_3\mathcal{L}'}{49\sigma_4}\nonumber\\
&-\frac{1772\alpha\beta\sigma_1\mathcal{W}_3\mathcal{L}\eta'}{49\sigma_4}-\frac{912\alpha\beta\eta\sigma_1\mathcal{L}\mathcal{W}_3'}{49\sigma_4}\\
 \mathit{f}_{-3}&\,=\,\frac{\mathit{f}^{(6)}}{720}+\frac{5}{72}\alpha\mathit{f}^{(4)}\mathcal{L}+\frac{1}{8}\alpha\mathit{f}^{(3)}\mathcal{L}'+\frac{40\gamma\sigma_1\mathcal{W}_4\mathit{f}''}{7\sigma_2}+\frac{34}{45}\alpha^2\mathcal{L}^2\mathit{f}''+\frac{13}{120}\alpha\mathit{f}''\mathcal{L}''+\frac{32\gamma\sigma_1\mathit{f}'\mathcal{W}_4'}{7\sigma_2}\nonumber\\
&+\frac{17}{360}\alpha\mathcal{L}^{(3)}\mathit{f}'+\frac{241}{180}\alpha^2\mathcal{L}\mathit{f}'\mathcal{L}'+\frac{200}{7}\beta^2\mathit{f}\sigma_1\mathcal{W}_3^2+\frac{8\gamma\mathit{f}\sigma_1\mathcal{W}_4''}{7\sigma_2}+\frac{1}{120}\alpha\mathit{f}\mathcal{L}^{(4)}+\alpha^3\mathit{f}\mathcal{L}^3\nonumber\\
&+\frac{23}{60}\alpha^2\mathit{f}\mathcal{L}\mathcal{L}''+\frac{3}{10}\alpha^2\mathit{f}\left(\mathcal{L}'\right)^2+\frac{176\alpha\gamma\mathit{f}\sigma_1\mathcal{W}_4\mathcal{L}}{7\sigma_2}+\frac{3840\beta\gamma\eta\sigma_1^2\mathcal{W}_3\mathcal{W}_4}{7\sigma_2\sigma_4}+\frac{41\beta\eta^{(4)}\sigma_1\mathcal{W}_3}{98\sigma_4}\nonumber
\end{align}
\begin{align}
&+\frac{99\beta\eta^{(3)}\sigma_1\mathcal{W}_3'}{98\sigma_4}+\frac{101\beta\sigma_1\eta''\mathcal{W}_3''}{98\sigma_4}+\frac{\beta\sigma_1\mathcal{W}_3{}^{(3)}\eta'}{2\sigma_4}+\frac{2\beta\eta\sigma_1\mathcal{W}_3{}^{(4)}}{21\sigma_4}+\frac{3}{10}\beta\sigma_2\mathcal{W}_3\chi''\nonumber\\
&+\frac{1}{6}\beta\sigma_2\chi'\mathcal{W}_3'+\frac{1}{30}\beta\sigma_2\chi\mathcal{W}_3''+\gamma\mathcal{W}_4\epsilon-\frac{16000\delta\sigma_1^2\mathcal{W}_5\eta''}{343\sigma_2\sigma_4^2}-\frac{320\delta\sigma_1^2\eta'\mathcal{W}_5'}{7\sigma_2\sigma_4^2}-\frac{640\delta\eta\sigma_1^2\mathcal{W}_5''}{49\sigma_2\sigma_4^2}\nonumber\\
&+\frac{80\delta\sigma_1\chi\mathcal{W}_5}{7\sigma_4}+\frac{1200\alpha^2\beta\eta\sigma_1\mathcal{W}_3\mathcal{L}^2}{49\sigma_4}+\frac{136\alpha\beta\eta\sigma_1\mathcal{W}_3\mathcal{L}''}{49\sigma_4}+\frac{468\alpha\beta\sigma_1\mathcal{W}_3\eta'\mathcal{L}'}{49\sigma_4}\nonumber\\
&+\frac{824\alpha\beta\eta\sigma_1\mathcal{W}_3'\mathcal{L}'}{147\sigma_4}+\frac{517\alpha\beta\sigma_1\mathcal{W}_3\mathcal{L}\eta''}{49\sigma_4}+\frac{599\alpha\beta\sigma_1\mathcal{L}\eta'\mathcal{W}_3'}{49\sigma_4}+\frac{596\alpha\beta\eta\sigma_1\mathcal{L}\mathcal{W}_3''}{147\sigma_4}\nonumber\\
&+\alpha\beta\sigma_2\chi\mathcal{W}_3\mathcal{L}-\frac{83200\alpha\delta\eta\sigma_1^2\mathcal{W}_5\mathcal{L}}{343\sigma_2\sigma_4^2}
\end{align}
\begin{align}
\eta_3&\,=\, -\eta '\\
\eta_2&\,=\, \frac{\eta ''}{2}+4 \alpha  \eta  \mathcal{L}\\
\eta_1&\,=\, -\frac{\eta ^{(3)}}{6}-\frac{4}{3} \alpha  \eta  \mathcal{L}'-\frac{11}{3} \alpha  \mathcal{L} \eta '\\
\eta_0&\,=\,3\beta\mathit{f}\sigma_4\mathcal{W}_3+\frac{\eta^{(4)}}{24}-\frac{480\gamma\eta\sigma_1\mathcal{W}_4}{7\sigma_2}+6\alpha^2\eta\mathcal{L}^2+\frac{1}{3}\alpha\eta\mathcal{L}''+\frac{5}{4}\alpha\eta'\mathcal{L}'+\frac{5}{3}\alpha\mathcal{L}\eta''\\
\eta_{-1}&\,=\,-\frac{13}{5}\beta\sigma_4\mathcal{W}_3\mathit{f}'-\frac{3}{5}\beta\mathit{f}\sigma_4\mathcal{W}_3'-\frac{\eta^{(5)}}{120}+\frac{48\gamma\sigma_1\mathcal{W}_4\eta'}{\sigma_2}+\frac{96\gamma\eta\sigma_1\mathcal{W}_4'}{7\sigma_2}-\frac{1}{15}\alpha\eta\mathcal{L}^{(3)}\nonumber\\
&-\frac{73}{15}\alpha^2\mathcal{L}^2\eta'-\frac{19}{60}\alpha\eta'\mathcal{L}''-\frac{56}{15}\alpha^2\eta\mathcal{L}\mathcal{L}'-\frac{7}{12}\alpha\eta''\mathcal{L}'-\frac{1}{2}\alpha\eta^{(3)}\mathcal{L}
\end{align}
\begin{align}
\eta_{-2}&\,=\,\frac{11}{10}\beta\sigma_4\mathcal{W}_3\mathit{f}''+\frac{8}{15}\beta\sigma_4\mathit{f}'\mathcal{W}_3'+\frac{1}{10}\beta\mathit{f}\sigma_4\mathcal{W}_3''-\frac{320\delta\mathit{f}\sigma_1\mathcal{W}_5}{7\sigma_2}+6\alpha\beta\mathit{f}\sigma_4\mathcal{W}_3\mathcal{L}+\frac{\eta^{(6)}}{720}\nonumber\\
&+\frac{320}{7}\beta^2\eta\sigma_1\mathcal{W}_3^2-\frac{792\gamma\sigma_1\mathcal{W}_4\eta''}{49\sigma_2}-\frac{72\gamma\sigma_1\eta'\mathcal{W}_4'}{7\sigma_2}-\frac{16\gamma\eta\sigma_1\mathcal{W}_4''}{7\sigma_2}+2\gamma\sigma_4\chi\mathcal{W}_4+\frac{1}{90}\alpha\eta\mathcal{L}^{(4)}\nonumber\\
&+4\alpha^3\eta\mathcal{L}^3+\frac{23}{360}\alpha\mathcal{L}^{(3)}\eta'+\frac{173}{90}\alpha^2\mathcal{L}^2\eta''+\frac{38}{45}\alpha^2\eta\mathcal{L}\mathcal{L}''+\frac{3}{20}\alpha\eta''\mathcal{L}''+\frac{277}{90}\alpha^2\mathcal{L}\eta'\mathcal{L}'\nonumber\\
&+\frac{28}{45}\alpha^2\eta\left(\mathcal{L}'\right)^2+\frac{13}{72}\alpha\eta^{(3)}\mathcal{L}'+\frac{1}{9}\alpha\eta^{(4)}\mathcal{L}-\frac{5440\alpha\gamma\eta\sigma_1\mathcal{W}_4\mathcal{L}}{49\sigma_2}
\end{align}
\begin{align}
\eta_{-3}&\,=\,-\frac{3}{10}\beta\mathit{f}^{(3)}\sigma_4\mathcal{W}_3-\frac{7}{30}\beta\sigma_4\mathit{f}''\mathcal{W}_3'-\frac{19}{210}\beta\sigma_4\mathit{f}'\mathcal{W}_3''+\frac{1280\delta\sigma_1\mathcal{W}_5\mathit{f}'}{49\sigma_2}-\frac{149}{35}\alpha\beta\sigma_4\mathcal{W}_3\mathcal{L}\mathit{f}'\nonumber\\
&-\frac{1}{70}\beta\mathit{f}\sigma_4\mathcal{W}_3{}^{(3)}+\frac{320\delta\mathit{f}\sigma_1\mathcal{W}_5'}{49\sigma_2}-\frac{12}{7}\alpha\beta\mathit{f}\sigma_4\mathcal{W}_3\mathcal{L}'-\frac{39}{35}\alpha\beta\mathit{f}\sigma_4\mathcal{L}\mathcal{W}_3'-\frac{\eta^{(7)}}{5040}-\frac{200}{7}\beta^2\sigma_1\mathcal{W}_3^2\eta'\nonumber\\
&-\frac{160}{7}\beta^2\eta\sigma_1\mathcal{W}_3\mathcal{W}_3'+\frac{1192\gamma\eta^{(3)}\sigma_1\mathcal{W}_4}{343\sigma_2}+\frac{1296\gamma\sigma_1\eta''\mathcal{W}_4'}{343\sigma_2}+\frac{88\gamma\sigma_1\eta'\mathcal{W}_4''}{49\sigma_2}+\frac{16\gamma\eta\sigma_1\mathcal{W}_4{}^{(3)}}{49\sigma_2}\nonumber\\
&-\frac{11}{7}\gamma\sigma_4\mathcal{W}_4\chi'-\frac{2}{7}\gamma\sigma_4\chi\mathcal{W}_4'-\frac{1}{630}\alpha\eta\mathcal{L}^{(5)}-\frac{3}{280}\alpha\mathcal{L}^{(4)}\eta'-\frac{93}{35}\alpha^3\mathcal{L}^3\eta'-\frac{47}{315}\alpha^2\eta\mathcal{L}\mathcal{L}^{(3)}\nonumber\\
&-\frac{11}{360}\alpha\mathcal{L}^{(3)}\eta''-\frac{22}{45}\alpha^2\eta^{(3)}\mathcal{L}^2-\frac{877\alpha^2\mathcal{L}\eta'\mathcal{L}''}{1260}-\frac{17}{360}\alpha\eta^{(3)}\mathcal{L}''-\frac{223}{180}\alpha^2\mathcal{L}\eta''\mathcal{L}'-\frac{37}{70}\alpha^2\eta'\left(\mathcal{L}'\right)^2\nonumber\\
&-\frac{1}{24}\alpha\eta^{(4)}\mathcal{L}'+\frac{8640\alpha\gamma\eta\sigma_1\mathcal{W}_4\mathcal{L}'}{343\sigma_2}-\frac{116}{35}\alpha^3\eta\mathcal{L}^2\mathcal{L}'-\frac{94}{315}\alpha^2\eta\mathcal{L}'\mathcal{L}''-\frac{7}{360}\alpha\eta^{(5)}\mathcal{L}\nonumber\\
&+\frac{21296\alpha\gamma\sigma_1\mathcal{W}_4\mathcal{L}\eta'}{343\sigma_2}+\frac{7456\alpha\gamma\eta\sigma_1\mathcal{L}\mathcal{W}_4'}{343\sigma_2}
\end{align}
\begin{align}
\eta_{-4}&\,=\,\mathcal{L}^4\eta\alpha^4+\frac{62}{63}\mathcal{L}\eta\left(\mathcal{L}'\right)^2\alpha^3+\frac{2747\mathcal{L}^2\mathcal{L}'\eta'\alpha^3}{1260}+\frac{197}{315}\mathcal{L}^2\eta\mathcal{L}''\alpha^3+\frac{256}{315}\mathcal{L}^3\eta''\alpha^3\nonumber\\
&+\frac{47\eta\left(\mathcal{L}''\right)^2\alpha^2}{1260}+3\mathit{f}\mathcal{L}^2\beta\mathcal{W}_3\sigma_4\alpha^2+\frac{517\mathcal{L}'\eta'\mathcal{L}''\alpha^2}{2016}+\frac{2227\left(\mathcal{L}'\right)^2\eta''\alpha^2}{10080}+\frac{88}{315}\mathcal{L}\mathcal{L}''\eta''\alpha^2\nonumber\\
&+\frac{47}{840}\eta\mathcal{L}'\mathcal{L}^{(3)}\alpha^2+\frac{613\mathcal{L}\eta'\mathcal{L}^{(3)}\alpha^2}{5040}+\frac{29}{90}\mathcal{L}\mathcal{L}'\eta^{(3)}\alpha^2+\frac{3}{140}\mathcal{L}\eta\mathcal{L}^{(4)}\alpha^2+\frac{4}{45}\mathcal{L}^2\eta^{(4)}\alpha^2\nonumber\\
&-\frac{2080\mathcal{L}^2\gamma\eta\mathcal{W}_4\sigma_1\alpha^2}{49\sigma_2}+\frac{2000}{49}\mathcal{L}\beta^2\eta\mathcal{W}_3^2\sigma_1\alpha+2\mathcal{L}\gamma\chi\mathcal{W}_4\sigma_4\alpha+\frac{253}{210}\beta\mathcal{W}_3\sigma_4\mathit{f}'\mathcal{L}'\alpha\nonumber\\
&-\frac{4792\gamma\mathcal{W}_4\sigma_1\mathcal{L}'\eta'\alpha}{343\sigma_2}+\frac{169}{210}\mathcal{L}\beta\sigma_4\mathit{f}'\mathcal{W}_3'\alpha+\frac{99}{280}\mathit{f}\beta\sigma_4\mathcal{L}'\mathcal{W}_3'\alpha-\frac{2012\gamma\eta\sigma_1\mathcal{L}'\mathcal{W}_4'\alpha}{343\sigma_2}\nonumber\\
&-\frac{4476\mathcal{L}\gamma\sigma_1\eta'\mathcal{W}_4'\alpha}{343\sigma_2}+\frac{146}{105}\mathcal{L}\beta\mathcal{W}_3\sigma_4\mathit{f}''\alpha+\frac{19}{56}\mathit{f}\beta\mathcal{W}_3\sigma_4\mathcal{L}''\alpha-\frac{1360\gamma\eta\mathcal{W}_4\sigma_1\mathcal{L}''\alpha}{343\sigma_2}\nonumber\\
&-\frac{5448\mathcal{L}\gamma\mathcal{W}_4\sigma_1\eta''\alpha}{343\sigma_2}+\frac{23}{140}\mathit{f}\mathcal{L}\beta\sigma_4\mathcal{W}_3''\alpha-\frac{1128\mathcal{L}\gamma\eta\sigma_1\mathcal{W}_4''\alpha}{343\sigma_2}+\frac{7}{720}\mathcal{L}^{(3)}\eta^{(3)}\alpha+\frac{13\eta''\mathcal{L}^{(4)}\alpha}{2520}\nonumber\\
&+\frac{1}{90}\mathcal{L}''\eta^{(4)}\alpha+\frac{31\eta'\mathcal{L}^{(5)}\alpha}{20160}+\frac{11\mathcal{L}'\eta^{(5)}\alpha}{1440}+\frac{\eta\mathcal{L}^{(6)}\alpha}{5040}+\frac{1}{360}\mathcal{L}\eta^{(6)}\alpha-\frac{32\mathit{f}\mathcal{L}\delta\mathcal{W}_5\sigma_1\alpha}{\sigma_2}\nonumber\\
&+\frac{20}{7}\beta^2\eta\sigma_1\left(\mathcal{W}_3'\right){}^2+\delta\epsilon\mathcal{W}_5+\frac{1}{2}\beta^2\chi\mathcal{W}_3^2\sigma_2\sigma_4+\frac{480\mathit{f}\beta\gamma\mathcal{W}_3\mathcal{W}_4\sigma_1\sigma_4}{7\sigma_2}+\frac{205}{14}\beta^2\mathcal{W}_3\sigma_1\eta'\mathcal{W}_3'\nonumber\\
&+\frac{13}{56}\gamma\sigma_4\chi'\mathcal{W}_4'-\frac{200\delta\sigma_1\mathit{f}'\mathcal{W}_5'}{49\sigma_2}-\frac{328\delta\mathcal{W}_5\sigma_1\mathit{f}''}{49\sigma_2}+\frac{815}{98}\beta^2\mathcal{W}_3^2\sigma_1\eta''+\frac{4}{7}\gamma\mathcal{W}_4\sigma_4\chi''\nonumber\\
&+\frac{30}{7}\beta^2\eta\mathcal{W}_3\sigma_1\mathcal{W}_3''+\frac{17}{420}\beta\sigma_4\mathit{f}''\mathcal{W}_3''+\frac{1}{28}\gamma\chi\sigma_4\mathcal{W}_4''-\frac{239\gamma\sigma_1\eta''\mathcal{W}_4''}{343\sigma_2}-\frac{40\mathit{f}\delta\sigma_1\mathcal{W}_5''}{49\sigma_2}\nonumber\\
&+\frac{1}{15}\beta\sigma_4\mathcal{W}_3'\mathit{f}^{(3)}-\frac{311\gamma\sigma_1\mathcal{W}_4'\eta^{(3)}}{343\sigma_2}+\frac{11}{840}\beta\sigma_4\mathit{f}'\mathcal{W}_3{}^{(3)}-\frac{13\gamma\sigma_1\eta'\mathcal{W}_4{}^{(3)}}{49\sigma_2}+\frac{7}{120}\beta\mathcal{W}_3\sigma_4\mathit{f}^{(4)}\nonumber\\
&-\frac{184\gamma\mathcal{W}_4\sigma_1\eta^{(4)}}{343\sigma_2}+\frac{1}{560}\mathit{f}\beta\sigma_4\mathcal{W}_3{}^{(4)}-\frac{2\gamma\eta\sigma_1\mathcal{W}_4{}^{(4)}}{49\sigma_2}+\frac{\eta^{(8)}}{40320}-\frac{21120\beta\delta\eta\mathcal{W}_3\mathcal{W}_5\sigma_1^2}{49\sigma_2\sigma_4}\nonumber\\
&+\frac{57600\gamma^2\eta\mathcal{W}_4^2\sigma_1^2}{49\sigma_2^2}\label{eta4}
\end{align}
\end{subequations}
which we will refer to as the twenty auxiliary equations.
\subsection{Ward idendities}\label{Wi}
In this Appendix we give in detail obtained spin 5 conformal Ward idendities of $AdS_3$ in Sec.\,\ref{spin5} as:
\begin{subequations}
\begin{align}
\delta _{\epsilon }\mathcal{W}_s&\,=\,s \mathcal{W}_s \epsilon '+\epsilon  \mathcal{W}_s'\,,(s=3,4,5)
\end{align}
\end{subequations}
define,\,that the each conformal field $ \mathcal{W}_s$ has the conformal spin s,\,with $s=3,4,5$.
\begin{subequations}
\begin{align}
\delta _{\chi }\mathcal{W}_3&\,=\,\frac{c}{360} \chi ^{(5)}+\frac{\chi '' \mathcal{L}'}{2}+\frac{1}{3} \chi ^{(3)} \mathcal{L}+\chi ' \left(\frac{32 \mathcal{L}^2}{5 c}+\frac{32 \mathcal{W}_4}{\sqrt{105}}+\frac{3 \mathcal{L}''}{10}\right)
+\chi  \left(\frac{32 \mathcal{L} \mathcal{L}'}{5 c}+\frac{16 \mathcal{W}_4'}{\sqrt{105}}+\frac{\mathcal{L}^{(3)}}{15}\right)
\end{align}
\end{subequations}
where $\chi$ is the gauge parameter related to the spin 3 charge in  $\mathfrak{sl}(5,\bR)$,\,which defines  spin 3 conformal Ward identity.
\begin{subequations}
\begin{align}
\delta_{\chi}\mathcal{W}_4&\,=\,
+\frac{4\chi^{(3)}\mathcal{W}_3}{\sqrt{105}}+\frac{4\chi''\mathcal{W}_3'}{\sqrt{105}}
+\chi' \left(\frac{416}{7c}\sqrt{\frac{3}{35}} \mathcal{L}\mathcal{W}_3
+\frac{4}{7}\sqrt{\frac{3}{35}} \mathcal{W}_3''+\frac{5\sqrt{15}\mathcal{W}_5}{7}\right)\nonumber\\
&+\chi\left(\frac{40\sqrt{\frac{15}{7}} \mathcal{W}_3\mathcal{L}'}{7c}+\frac{144\sqrt{\frac{3}{35}} \mathcal{L}\mathcal{W}_3'}{7c}+\frac{2}{7}\sqrt{15}\mathcal{W}_5'+\frac{2\mathcal{W}_3{}^{(3)}}{7\sqrt{105}}\right)
\end{align}
\end{subequations}
\begin{subequations}
\begin{align}
\delta_{\mathit{f}}\mathcal{W}_4&\,=\,
 \frac{c \mathit{f}^{(7)}}{20160}
+\frac{1}{60} \mathit{f}^{(5)} \mathcal{L}
+\frac{1}{24} \mathit{f}^{(4)} \mathcal{L}'
+\mathit{f}^{(3)} \left(\frac{7 \mathcal{L}^2}{5 c}+\frac{1}{2} \sqrt{\frac{3}{35}}\mathcal{W}_4+\frac{\mathcal{L}''}{20}\right)\nonumber\\
&+\mathit{f}'' \left(\frac{21 \mathcal{L} \mathcal{L}'}{5 c}+\frac{3}{4} \sqrt{\frac{3}{35}}\mathcal{W}_4'+\frac{\mathcal{L}^{(3)}}{30}\right)
+\mathit{f}'\Bigg(\frac{864\mathcal{L}^3}{35c^2}+\frac{45\mathcal{W}_3^2}{2c}+\frac{88\mathcal{L}\mathcal{L}''}{35c}+\frac{59\left(\mathcal{L}'\right)^2}{28c}\nonumber\\
&+\frac{4\mathcal{W}_4\sqrt{\frac{21}{5}} \mathcal{L}}{c}+\frac{1}{4}\sqrt{\frac{5}{21}} \mathcal{W}_4''+\frac{\mathcal{L}^{(4)}}{84}\Bigg)
+\mathit{f} \Bigg(\frac{1296 \mathcal{L}^2 \mathcal{L}'}{35 c^2}+\frac{45 \mathcal{W}_3 \mathcal{W}_3'}{2 c}+\frac{39 \mathcal{L}^{(3)} \mathcal{L}}{70 c}\nonumber\\
&+\frac{2 \sqrt{\frac{21}{5}} \mathcal{W}_4 \mathcal{L}'}{c}+\frac{177\mathcal{L}' \mathcal{L}''}{140 c}+\frac{2 \sqrt{\frac{21}{5}}\mathcal{L} \mathcal{W}_4'}{c}+\frac{\mathcal{W}_4{}^{(3)}}{4\sqrt{105}}+\frac{\mathcal{L}^{(5)}}{560}\Bigg)
\end{align}
\end{subequations}
where $\mathit{f}$ is the gauge parameter related to the spin 4 charge in $\mathfrak{sl}(5,\bR)$,\,which defines  spin 4 conformal Ward identity.
\begin{subequations}
\begin{align}
\delta_{\chi}\mathcal{W}_5&\,=\,
\frac{2}{7}\sqrt{\frac{5}{3}}\chi^{(3)}\mathcal{W}_4
+\frac{1}{14}\sqrt{15}\chi''\mathcal{W}_4'
+\chi'\left(\frac{24\mathcal{W}_3^2}{\sqrt{7}c}+\frac{32\sqrt{15}\mathcal{W}_4\mathcal{L}}{7c}+\frac{1}{14}\sqrt{\frac{5}{3}}\mathcal{W}_4''\right)\nonumber\\
&+\chi\left(\frac{16\mathcal{W}_3\mathcal{W}_3'}{\sqrt{7}c}+\frac{32\mathcal{W}_4\mathcal{L}'}{\sqrt{15}c}+\frac{128\mathcal{L}\mathcal{W}_4'}{7\sqrt{15}c}+\frac{\mathcal{W}_4{}^{(3)}}{21\sqrt{15}}\right)
\end{align}
\end{subequations}
\begin{subequations}
\begin{align}
\delta_{\mathit{f}}\mathcal{W}_5&\,=\,
 \frac{1}{56} \sqrt{\frac{3}{5}} \mathit{f}^{(5)} \mathcal{W}_3
+\frac{1}{56} \sqrt{\frac{5}{3}} \mathit{f}^{(4)} \mathcal{W}_3'
+\mathit{f}^{(3)}\left(\frac{33 \sqrt{15} \mathcal{W}_3 \mathcal{L}}{49 c}+\frac{1}{196} \sqrt{15} \mathcal{W}_3''-\frac{11}{28} \sqrt{\frac{5}{21}} \mathcal{W}_5\right)\nonumber\\
&+\mathit{f}'' \left(\frac{48 \sqrt{15} \mathcal{W}_3 \mathcal{L}'}{49 c}+\frac{34 \sqrt{15} \mathcal{L} \mathcal{W}_3'}{49 c}-\frac{11}{14} \sqrt{\frac{3}{35}} \mathcal{W}_5'+\frac{1}{392} \sqrt{15} \mathcal{W}_3{}^{(3)}\right)\nonumber\\
&+\mathit{f}'\Bigg(\frac{4896}{49c^2}\sqrt{\frac{3}{5}}\mathcal{W}_3\mathcal{L}^2+\frac{74\mathcal{W}_3\mathcal{W}_4}{\sqrt{7}c}+\frac{563}{196c}\sqrt{\frac{3}{5}}\mathcal{W}_3
\mathcal{L}''+\frac{397}{196c}\sqrt{\frac{5}{3}}\mathcal{W}_3'\mathcal{L}'+\frac{89}{98c}\sqrt{\frac{5}{3}}\mathcal{L}\mathcal{W}_3''\nonumber\\
&-\frac{58}{7c}\sqrt{\frac{15}{7}}\mathcal{W}_5\mathcal{L}-\frac{1}{14}\sqrt{\frac{15}{7}}\mathcal{W}_5''+\frac{5\sqrt{\frac{5}{3}}\mathcal{W}_3{}^{(4)}}{2352}\Bigg)
+\mathit{f}\Bigg(\frac{1728}{49c^2}\sqrt{\frac{3}{5}}\mathcal{L}^2\mathcal{W}_3'+\mathcal{W}_3\frac{4752}{49c^2}\sqrt{\frac{3}{5}}\mathcal{L}\mathcal{L}'\nonumber\\
&+\frac{6\sqrt{7}\mathcal{W}_4\mathcal{W}_3'}{c}+\frac{24\mathcal{W}_3\mathcal{W}_4'}{\sqrt{7}c}+\frac{123}{196c}\sqrt{\frac{3}{5}}\mathcal{W}_3\mathcal{L}^{(3)}+\frac{97\sqrt{\frac{3}{5}}\mathcal{W}_3'\mathcal{L}''}{98c}
+\frac{29\sqrt{15}\mathcal{W}_3''\mathcal{L}'}{196c}\nonumber\\
&-\frac{4}{c}\sqrt{\frac{15}{7}}\mathcal{W}_5\mathcal{L}'-\frac{118}{7c}\sqrt{\frac{3}{35}}\mathcal{L}\mathcal{W}_5'+\frac{5\sqrt{15}
\mathcal{W}_3{}^{(3)}\mathcal{L}}{98c}-\frac{1}{28}\sqrt{\frac{5}{21}}\mathcal{W}_5{}^{(3)}+\frac{\mathcal{W}_3{}^{(5)}}{784\sqrt{15}}\Bigg)
\end{align}
\end{subequations}
\begin{subequations}
\begin{align}
\delta_{\eta}\mathcal{W}_5&\,=\,
 \frac{c \eta ^{(9)}}{1814400}
 +\frac{\eta ^{(7)} \mathcal{L}}{2520}
 +\frac{1}{720} \eta ^{(6)} \mathcal{L}'\nonumber\\
&+\eta ^{(5)} \left(\frac{13 \mathcal{L}^2}{150 c}-\frac{11 \mathcal{W}_4}{140 \sqrt{105}}+\frac{\mathcal{L}''}{400}\right)
 +\eta ^{(4)} \left(\frac{13 \mathcal{L} \mathcal{L}'}{30 c}-\frac{11 \mathcal{W}_4'}{56 \sqrt{105}}+\frac{\mathcal{L}^{(3)}}{360}\right)\nonumber\\
&+\eta^{(3)}\Bigg(\frac{656\mathcal{L}^3}{105c^2}+\frac{181\mathcal{W}_3^2}{84c}+\frac{109\mathcal{L}\mathcal{L}''}{210c}+\frac{73\left(\mathcal{L}'\right)^2}{168c}-\frac{374\mathcal{W}_4\mathcal{L}}{21\sqrt{105}c}-\frac{11}{252}\sqrt{\frac{5}{21}}\mathcal{W}_4''\nonumber\\
&+\frac{\mathcal{L}^{(4)}}{504}\Bigg)
+\eta''\Bigg(\frac{984\mathcal{L}^2\mathcal{L}'}{35c^2}+\frac{181\mathcal{W}_3\mathcal{W}_3'}{28c}+\frac{29\mathcal{L}^{(3)}\mathcal{L}}{84c}
-\frac{187\mathcal{W}_4\mathcal{L}'}{7\sqrt{105}c}+\frac{219\mathcal{L}'\mathcal{L}''}{280c}\nonumber\\
&-\frac{187\mathcal{L}\mathcal{W}_4'}{7\sqrt{105}c}-\frac{11\mathcal{W}_4{}^{(3)}}{84\sqrt{105}}+\frac{\mathcal{L}^{(5)}}{1120}\Bigg)
+\eta\Bigg(\frac{36864\mathcal{L}^3\mathcal{L}'}{175c^3}-\frac{11008\mathcal{L}^2\mathcal{W}_4'}{35\sqrt{105}c^2}-\frac{22016\mathcal{W}_4\mathcal{L}\mathcal{L}'}{35\sqrt{105}c^2}\nonumber\
\end{align}
\end{subequations}
\begin{subequations}
\begin{align}
&+\frac{496\mathcal{W}_3^2
\mathcal{L}'}{7c^2}+\frac{496\left(\mathcal{L}'\right)^3}{105c^2}+\frac{1952\mathcal{L}^{(3)}\mathcal{L}^2}{525c^2}+\frac{8864\mathcal{L}\mathcal{L}'\mathcal{L}''}{525c^2}+\frac{992\mathcal{W}_3\mathcal{L}
\mathcal{W}_3'}{7c^2}+\frac{5\mathcal{W}_3'\mathcal{W}_3''}{3c}\nonumber\\
&-\frac{30\mathcal{W}_5\mathcal{W}_3'}{\sqrt{7}c}+\frac{448\mathcal{W}_4\mathcal{W}_4'}{15c}-\frac{30\mathcal{W}_3\mathcal{W}_5'}{\sqrt{7}
c}+\frac{19\mathcal{W}_3\mathcal{W}_3{}^{(3)}}{21c}+\frac{29\mathcal{L}^{(5)}\mathcal{L}}{1575c}-\frac{368\mathcal{W}_4\mathcal{L}^{(3)}}{105\sqrt{105}c}-\frac{281\mathcal{W}_4'\mathcal{L}''}{35\sqrt{105}c}\nonumber\\
&-\frac{157\mathcal{W}_4''\mathcal{L}'}{21\sqrt{105}c}+\frac{13\mathcal{L}^{(4)}\mathcal{L}'}{210c}+\frac{47\mathcal{L}^{(3)}\mathcal{L}''}{450c}-\frac{62\mathcal{W}_4{}^{(3)}\mathcal{L}}{21\sqrt{105}
c}-\frac{\mathcal{W}_4{}^{(5)}}{180\sqrt{105}}+\frac{\mathcal{L}^{(7)}}{37800}\Bigg)
\end{align}
\end{subequations}
where $\eta$ is the gauge parameter related to the spin 5 charge in $\mathfrak{sl}(5,\bR)$,\,which defines  spin 5 conformal Ward identities.
\newpage
\section{Classical $\cW_5$ symmetry OPEs from the conformal Ward identities}\label{appE}
In this part we will give in detail obtained the classical $\cW_5$ symmetry OPEs from the conformal Ward identities in Appendix \ref{Wi}.
\begin{subequations}
\begin{align}
\mathcal{L}(&z_1)\mathcal{L}(z_2)\,\sim \,{{c\over 2}\over{z_{12}^{4}}}\,+\, {2\,\mathcal{L}\over{z_{12}^{2}}}\, + \,{ \mathcal{L}'\over{z_{12}}}\\
\mathcal{L}(&z_1)\mathcal{W}_s(z_2)\,\sim \,{s\,\mathcal{W}_s\over{z_{12}^{2}}}\, + \,{ \mathcal{W}_s'\over{z_{12}}}\,\,{~,~~~~}(s\,=\,3,4,5)\\
\mathcal{W}_3(&z_1)\mathcal{W}_3(z_2)\,\sim \,\nonumber
{{c\over 3}\over{z_{12}^{6}}}\,+\, {2\,\mathcal{L}\over{z_{12}^{4}}}\, + \,{ \mathcal{L}'\over{z_{12}^{3}}}
\,+\,\frac{1}{z_{12}^2}\Big(\frac {32 \mathcal {L}^2} {5 c} + \frac {32 \mathcal {W} _ 4} {\sqrt {105}} + \frac {3 \mathcal {L}''} {10}\Big)\nonumber
\,+\, \frac{1}{z_{12}}\Big(\frac{32 \mathcal{L} \mathcal{L}'}{5 c}+\frac{16 \mathcal{W}_4'}{\sqrt{105}}+\frac{\mathcal{L}{}^{(3)}}{15}\Big)\\
 \mathcal{W}_3(&z_1)\mathcal{W}_4(z_2)\,\sim \,\nonumber
\frac{8}{\sqrt{105}}\left(\frac{3 \mathcal{W}_3}{z_{12}^4}+\frac{\mathcal{W}_3'}{z_{12}^3}\right)
 \,+\,\frac{1}{z_{12}^2}
 \Big(\frac{416}{7c} \sqrt{\frac{3}{35}} \mathcal{L} \mathcal{W}_3
  +\frac{4}{7} \sqrt{\frac{3}{35}} \mathcal{W}_3''+\frac{5}{7} \sqrt{15} \mathcal{W}_5\Big)\nonumber\\
&\,+\,\frac{1}{z_{12}}\Big(\frac{40}{7 c}\sqrt{\frac{15}{7}} \mathcal{W}_3 \mathcal{L}'+\frac{144}{7 c}\sqrt{\frac{3}{35}} \mathcal{L} \mathcal{W}_3'+\frac{2}{7} \sqrt{15} \mathcal{W}_5'+\frac{2 \mathcal{W}_3{}^{(3)}}{7
\sqrt{105}}\Big)\\
\mathcal{W}_4(&z_1)\mathcal{W}_4(z_2)\,\sim \,
{{c\over 4}\over{z_{12}^{8}}}\,+\, {2\,\mathcal{L}\over{z_{12}^{6}}}\, + \,{ \mathcal{L}'\over{z_{12}^{5}}}
\,+\, \frac{1}{z_{12}^4}\Big(3\sqrt{\frac{3}{35}}\mathcal{W}_4\,+\,\frac{42 }{5 c}\mathcal{L}^2\,+\,{3 \over 10}\mathcal{L}''\Big)\nonumber
\,+\, \frac{1}{z_{12}^3}\Big(\frac{3}{2}\sqrt{\frac{3}{35}}\mathcal{W}_4'\,+\,\frac{42 }{5 c}\mathcal{L} \mathcal{L}'\,+\,\frac{\mathcal{L}{}^{(3)}}{15}\Big)\nonumber \\
&\,+\, \frac{1}{z_{12}^2}\Big(\frac{864}{35 c^2}\mathcal{L}^3+\frac{59}{28 c}\left(\mathcal{L}'\right){}^2+\frac{88}{35 c}\mathcal{L} \mathcal{L}''+\frac{45}{2
   c}\mathcal{W}_3^2+\frac{4}{c}\sqrt{\frac{21}{5}}\mathcal{L} \mathcal{W}_4+\frac{1}{4}\sqrt{\frac{5}{21}}\mathcal{W}_4''+\frac{\mathcal{L}{}^{(4)}}{84}\Big)\nonumber \\
&\,+\, \frac{1}{z_{12}}\Big( \frac{1296}{35 c^2}\mathcal{L}^3\mathcal{L}'+\frac{177}{140 c}\mathcal{L}'\mathcal{L}''+\frac{45}{2
   c}\mathcal{W}_3\mathcal{W}_3'+\frac{2}{c}\sqrt{\frac{21}{5}}\mathcal{L}\mathcal{W}_4'+\frac{2}{c}\sqrt{\frac{21}{5}}\mathcal{W}_4\mathcal{L}'\nonumber
+\frac{39}{70 c}\mathcal{L}\mathcal{L}{}^{(3)} +\frac{\mathcal{W}_4{}^{(3)}}{4 \sqrt{105}}+\frac{\mathcal{L}{}^{(5)}}{560} \Big)\\
\mathcal{W}_3(&z_1)\mathcal{W}_5(z_2)\,\sim \,
\frac{\sqrt{15}}{7}\left(\frac{4 \mathcal{W}_4}{z_{12}^4}+\frac{\mathcal{W}_4'}{z_{12}^3}\right)
\,+\,\frac {1} {z_ {12}^2}\Big (\frac {24\mathcal {W} _ 3^2} {\sqrt {7} c} + \frac {32\sqrt {15}\mathcal {L}\mathcal {W} _ 4} {7 c} + \frac {1} {14}\sqrt {\frac {5} {3}}\mathcal {W} _ 4''\Big)\nonumber\\
&\,+\, \frac {1} {z_ {12}}\Big (\frac {32\mathcal {W} _ 4\mathcal {L}'} {\sqrt {15} c} + \frac {16\mathcal {W} _ 3\mathcal {W} _ 3'}{\sqrt {7} c} + \frac {128\mathcal {L}\mathcal {W} _ 4'} {7\sqrt{15} c} + \frac {\mathcal {W} _ 4 {}^{(3)}} {21\sqrt {15}}\Big)
\end{align}
\begin{align}
\mathcal{W}_4(&z_1)\mathcal{W}_5(z_2)\,\sim \,
\frac{\sqrt{15}}{7}\left(+\frac{3 \mathcal{W}_3}{z_{12}^6}+\frac{\mathcal{W}_3'}{z_{12}^5}\right)
\, +\, \frac {1} {z_ {12}^4}\Big (\frac {198\sqrt {15}\mathcal {L}\mathcal {W} _ 3} {49 c} + \frac {3} {98}\sqrt {15}\mathcal {W} _ 3''- \frac {11} {14}\sqrt {\frac {15} {7}}\mathcal {W} _ 5\Big)\nonumber\\
&\, +\, \frac {1} {z_ {12}^3}\Big (\frac {96\sqrt {15}\mathcal {W} _ 3\mathcal {L}'} {49 c} + \frac {68\sqrt {15}\mathcal {L}\mathcal {W} _ 3'} {49 c} - \frac {11} {7}\sqrt {\frac {3}{35}}\mathcal {W} _ 5' + \frac {1} {196}\sqrt {15}\mathcal {W} _ 3 {}^{(3)}\Big)\nonumber\\
&\, +\, \frac {1} {z_ {12}^2}\Big (\frac {4896\sqrt {\frac {3}{5}}\mathcal {W} _ 3\mathcal {L}^2} {49 c^2} + \frac {397\sqrt {\frac {5} {3}}\mathcal {L}'\mathcal {W} _ 3'} {196 c} + \frac{89\sqrt {\frac {5} {3}}\mathcal {L}\mathcal {W} _ 3''} {98 c} +\frac {563\sqrt {\frac {3} {5}}\mathcal {W} _ 3\mathcal {L}''}{196 c}\nonumber\\
& - \frac {58\sqrt {\frac {15} {7}}\mathcal {W} _ 5\mathcal {L}} {7 c} + \frac {74\mathcal {W} _ 3\mathcal {W} _ 4} {\sqrt {7} c} - \frac {1} {14}\sqrt {\frac {15} {7}}\mathcal {W} _ 5'' + \frac {5\sqrt {\frac {5} {3}}\mathcal {W} _ 3 {}^{(4)}} {2352}\Big)\nonumber
\end{align}
\begin{align}
&+\, \frac {1} {z_ {12}}\Big (\frac {1728\sqrt {\frac {3} {5}}\mathcal {L}^2\mathcal {W} _ 3'} {49 c^2} + \frac {4752\sqrt {\frac {3} {5}}\mathcal {W} _ 3\mathcal {L}\mathcal {L}'} {49 c^2} + \frac {97\sqrt {\frac {3} {5}}\mathcal {W} _ 3'\mathcal {L}''} {98 c} + \frac {29\sqrt {15}\mathcal {L}'\mathcal {W} _ 3''}{196 c}\nonumber\\
& - \frac {118\sqrt {\frac {3} {35}}\mathcal {L}\mathcal {W} _ 5'} {7 c} - \frac {4\sqrt {\frac {15} {7}}\mathcal {W} _ 5\mathcal {L}'} {c} + \frac {6\sqrt {7}\mathcal {W} _ 4\mathcal {W} _ 3'} {c} + \frac {24\mathcal {W} _ 3\mathcal {W} _ 4'} {\sqrt {7} c} + \frac {5\sqrt {15}\mathcal {W} _ 3 {}^{(3)}\mathcal {L}} {98 c}\nonumber\\
& + \frac {123\sqrt {\frac {3} {5}}\mathcal {W} _ 3\mathcal {L} {}^{(3)}} {196 c} - \frac {1} {28}\sqrt {\frac {5} {21}}\mathcal {W} _ 5 {}^{(3)} + \frac {\mathcal {W} _ 3 {}^{(5)}} {784\sqrt {15}}\Big)
\end{align}
\begin{align}
\mathcal{W}_5(&z_1)\mathcal{W}_5(z_2)\,\sim\,\nonumber
{{c\over5}\over{z_{12}^{10}}}\,+\,{2\,\mathcal{L}\over{z_{12}^{8}}}\,+\,{\mathcal{L}'\over{z_{12}^{7}}}\nonumber\\
&\,+\,\frac{1}{z_{12}^6}\Big(\frac{52\mathcal{L}^2}{5c}-\frac{22}{7}\sqrt{\frac{3}{35}}\mathcal{W}_4+\frac{3\mathcal{L}''}{10}\Big)\nonumber\\
& \,+\,\frac{1}{z_{12}^5}\Big(\frac{52\mathcal{L}\mathcal{L}'}{5c}-\frac{11}{7}\sqrt{\frac{3}{35}}\mathcal{W}_4'+\frac{\mathcal{L}{}^{(3)}}{15}\Big)\nonumber\\
&\,+\,\frac{1}{z_{12}^4}\Big(\frac{1312\mathcal{L}^3}{35c^2}+\frac{73\left(\mathcal{L}'\right){}^2}{28c}+\frac{109\mathcal{L}\mathcal{L}''}{35c}-\frac{748\mathcal{W}_4
\mathcal{L}}{7\sqrt{105}c}+\frac{181\mathcal{W}_3^2}{14c}-\frac{11}{42}\sqrt{\frac{5}{21}}\mathcal{W}_4''+\frac{\mathcal{L}{}^{(4)}}{84}\Big)\nonumber\\
&\,+\,\frac{1}{z_{12}^3}\Big(\frac{1968\mathcal{L}^2\mathcal{L}'}{35c^2}+\frac{219\mathcal{L}'\mathcal{L}''}{140c}-\frac{374\mathcal{L}\mathcal{W}_4'}{7\sqrt{105}c}-\frac{374\mathcal{W}_4\mathcal{L}'}{7\sqrt{105}c}+\frac{181\mathcal{W}_3\mathcal{W}_3'}{14c}\nonumber
 \,+\,\frac{29\mathcal{L}{}^{(3)}\mathcal{L}}{42c}-\frac{11\mathcal{W}_4{}^{(3)}}{42\sqrt{105}}+\frac{\mathcal{L}{}^{(5)}}{560}\Big)\nonumber\\
&\,+\,\frac{1}{z_{12}^2}\Big(\frac{18432\mathcal{L}^4}{175c^3}+\frac{592\mathcal{L}\left(\mathcal{L}'\right){}^2}{21c^2}+\frac{8824\mathcal{L}^2\mathcal{L}''}{525c^2}-\frac{22016\mathcal{W}_4\mathcal{L}^2}{35\sqrt
{105}c^2}\nonumber+\frac{992\mathcal{W}_3^2\mathcal{L}}{7c^2}\nonumber\\
&+\frac{35\left(\mathcal{W}_3'\right){}^2}{12c}-\frac{188\mathcal{L}'\mathcal{W}_4'}{7\sqrt{105}c}+\frac{25\mathcal{L}{}^{(3)}\mathcal{L}'}{72c}-\frac{311\mathcal{L}\mathcal{W}_4''}{21\sqrt{105}c}+\frac{47
\left(\mathcal{L}''\right){}^2}{200c}-\frac{557\mathcal{W}_4\mathcal{L}''}{35\sqrt{105}c}\nonumber\\
&+\frac{111\mathcal{W}_3\mathcal{W}_3''}{28c}+\frac{31\mathcal{L}{}^{(4)}\mathcal{L}}{252c}+\frac{448\mathcal{W}_4^2}{15
c}-\frac{60\mathcal{W}_3\mathcal{W}_5}{\sqrt{7}c}-\frac{\mathcal{W}_4{}^{(4)}}{24\sqrt{105}}+\frac{\mathcal{L}{}^{(6)}}{4320}\Big)\nonumber\\
&\,+\,\frac{1}{z_{12}}\Big(\frac{36864\mathcal{L}^3\mathcal{L}'}{175c^3}-\frac{11008\mathcal{L}^2\mathcal{W}_4'}{35\sqrt{105}c^2}-\frac{22016\mathcal{W}_4\mathcal{L}\mathcal{L}'}{35\sqrt{105}c^2}+\frac{496\mathcal{W}_3^2\mathcal{L}'}{7c^2}+\frac{496\left(\mathcal{L}'\right)^3}{105c^2}\nonumber\\
&+\frac{1952\mathcal{L}^{(3)}\mathcal{L}^2}{525c^2}+\frac{8864\mathcal{L}\mathcal{L}'\mathcal{L}''}{525c^2}+\frac{992\mathcal{W}_3\mathcal{L}\mathcal{W}_3'}{7c^2}+\frac{5\mathcal{W}_3'\mathcal{W}_3''}{3c}-\frac{30\mathcal{W}_5\mathcal{W}_3'}{\sqrt{7}c}
+\frac{448\mathcal{W}_4\mathcal{W}_4'}{15c}\nonumber\\
&-\frac{30\mathcal{W}_3\mathcal{W}_5'}{\sqrt{7}c}+\frac{19\mathcal{W}_3\mathcal{W}_3{}^{(3)}}{21c}+\frac{29\mathcal{L}^{(5)}\mathcal{L}}{1575c}-\frac{368\mathcal{W}_4\mathcal{L}^{(3)}}{105\sqrt{105}c}-\frac{281\mathcal{W}_4'\mathcal{L}''}{35\sqrt{105}c}-\frac{157\mathcal{W}_4''\mathcal{L}'}{21\sqrt{105}c}\nonumber\\
&+\frac{13\mathcal{L}^{(4)}\mathcal{L}'}{210c}+\frac{47\mathcal{L}^{(3)}\mathcal{L}''}{450c}-\frac{62\mathcal{W}_4{}^{(3)}\mathcal{L}}{21\sqrt{105}c}-\frac{\mathcal{W}_4{}^{(5)}}{180\sqrt{105}}+\frac{\mathcal{L}^{(7)}}{37800}
\Big)
\end{align}
\end{subequations}

\section{Chemical Potentials for $\mathfrak{sl}(5,\bR) \oplus \mathfrak{sl}(5,\bR)$ Chern\,-\,Simons Theory }\label{cp5}
The final form of the spatial and temporal connection with the spin 5 chemical potential in Sec.\,\ref{sl5}  is given by
\begin{align}
a_{\phi}(t,\phi)&=
  \Lt_{1}
+ \frac{6}{c} \cL \Lt_{-1}
+ \frac{15}{c} \cW_3 \Wt_{-2}^{(3)}
+ \frac{28}{c} \cW_4 \Wt_{-3}^{(4)}
+ \frac{45}{c} \cW_5 \Wt_{-4}^{(5)}
 , \nonumber \\
a_{t}(t,\phi)&=
  \mu_{2}  \Lt_{1}
+ \mu_{3}  \Wt_{2}^{(3)}
+ \mu_{4}  \Wt_{3}^{(4)}
+ \mu_{5}  \Wt_{4}^{(5)}\nonumber\\
&+\sum_{i=-1}^{0} \nu_{2}^{(i)}  \Lt_{i}
+ \sum_{i=-2}^{1} \nu_{3}^{(i)}  \Wt_{i}^{(3)}
+ \sum_{i=-3}^{2} \nu_{4}^{(i)}  \Wt_{i}^{(4)}
+ \sum_{i=-4}^{3} \nu_{5}^{(i)}  \Wt_{i}^{(5)}.\nonumber \\
&=a_{t}^{( \mu_{2})}+a_{t}^{( \mu_{3})}+a_{t}^{( \mu_{4})}+a_{t}^{( \mu_{5})}
\end{align}
where
\begin{subequations}
  \begin{align}
&a_{t}^{( \mu_{2})}=
\mu_2 \bigg(
  \Lt_{1}
+ \frac{6}{c} \cL \Lt_{-1}
+ \frac{15}{c} \cW_3 \Wt_{-2}^{(3)}
+ \frac{28}{c} \cW_4 \Wt_{-3}^{(4)}
+ \frac{45}{c} \cW_5 \Wt_{-4}^{(5)}
\bigg)
- \mu_2' \Lt_{0}
+ \frac{1}{2} \mu _2''\Lt_{-1}
  \end{align}
  \begin{align}
&a_{t}^{( \mu_{3})}=
\mu_3\bigg(\Wt_2^{(3)}+\frac{12}{c}\mathcal{W}_3\Lt_{-1}\nonumber\\
&+\bigg(\frac{36}{c^2}\cL^2+\frac{16}{c}\sqrt{\frac{15}{7}}\mathcal{W}_4+\frac{1}{c}\cL'' \bigg)\Wt_{-2}^{(3)}+\bigg(\frac{1800}{7 \sqrt{7}
   c^2}\mathcal{W}_3^2+\frac{48 \sqrt{15}}{c^2}\cL\mathcal{W}_4+\frac{\sqrt{15}}{7
   c}\mathcal{W}_4''\bigg)\Wt_{-4}^{(5)}\nonumber\\
&+\bigg(\frac{96}{c^2}\sqrt{\frac{15}{7}}\cL\mathcal{W}_3+\frac{8}{\sqrt{105} c}\mathcal{W}_3''+\frac{8
   \sqrt{15}}{c}\mathcal{W}_5\bigg)\Wt_{-3}^{(4)}-\frac{8 \sqrt{15}}{7
   c}\mathcal{W}_4'\Wt_{-3}^{(5)}-\frac{16}{c}\sqrt{\frac{3}{35}}\mathcal{W}_3'\Wt_{-2}^{(4)}\nonumber\\
&+\frac{8\sqrt{15}}{c}\mathcal{W}_4\Wt_{-2}^{(5)}+\frac{16}{c}\sqrt{\frac{15}{7}}\mathcal{W}_3\Wt_{-1}^{(4)}-\frac{4}{c}\cL'\Wt_{-1}^{(3)}+\frac{12}{c}
   \cL\Wt_0^{(3)} \bigg)
 +\mu_3'\bigg(\frac{13 \sqrt{15}}{14
   c}\mathcal{W}_4'\Wt_{-4}^{(5)}\nonumber\\ &+\frac{8}{c}\sqrt{\frac{5}{21}}\mathcal{W}_3'\Wt_{-3}^{(4)}-\frac{44 \sqrt{15}}{7
   c}\mathcal{W}_4\Wt_{-3}^{(5)}-\frac{64}{c}\sqrt{\frac{3}{35}}\mathcal{W}_3\Wt_{-2}^{(4)}+\frac{7}{2
   c}\cL'\Wt_{-2}^{(3)}-\frac{10}{c}\cL\Wt_{-1}^{(3)}-\Wt_1^{(3)}\bigg)\nonumber\\
&+\mu _3''\bigg(\frac{16 \sqrt{15}}{7
   c}\mathcal{W}_4\Wt_{-4}^{(5)}+\frac{24}{c}\sqrt{\frac{3}{35}}\mathcal{W}_3\Wt_{-3}^{(4)}+\frac{4}{c}\cL\Wt_{-2}^{(3)}+\frac{1}{2}\Wt_0^{(3)}\bigg)
   -\frac{\mu _3^{(3)}}{6}\Wt_{-1}^{(3)}
   +\frac{\mu _3^{(4)}}{24}\Wt_{-2}^{(3)}
 \end{align}
 \begin{align}
&a_{t}^{( \mu_{4})}=
 \mu_{4}\bigg(\Wt_3^{(4)}+\frac{18}{c}\mathcal{W}_4\Lt_{-1} \nonumber\\
&+\bigg(\frac{108}{c^2}\cL^2+\frac{3}{2 c}\cL''+\frac{2 \sqrt{105}}{c}\cW_4\bigg)\Wt_{-1}^{(4)}
+\bigg(\frac{3 \sqrt{15}}{784 c}\cW_3^{(4)}+\frac{1620 \sqrt{15}}{7 c^3}\cL^2\cW_3+\frac{900}{\sqrt{7} c^2}\cW_3\cW_4\nonumber\\
&+\frac{207 \sqrt{15}}{98 c^2}\cL\cW_3''+\frac{855 \sqrt{15}}{196 c^2}\cW_3\cL''+\frac{891 \sqrt{15}}{196 c^2}\cL'\cW_3'-\frac{18 \sqrt{105}}{c^2}\cL\cW_5-\frac{15}{28 c}\sqrt{\frac{15}{7}}\cW_5''\bigg)\Wt_{-4}^{(5)}\nonumber\\
&-\bigg(\frac{3}{10 c}\cL^{(3)}+\frac{324}{5 c^2}\cL\cL'+\frac{2}{c}\sqrt{\frac{21}{5}}\cW_4'\bigg)\Wt_{-2}^{(4)}
+\bigg(-\frac{3 \sqrt{15}}{98 c}\cW_3^{(3)}-\frac{702 \sqrt{15}}{49 c^2}\cL\cW_3'-\frac{1080 \sqrt{15}}{49 c^2}\cW_3\cL'\nonumber\\
&+\frac{30}{7 c}\sqrt{\frac{15}{7}}\cW_5'\bigg)\Wt_{-3}^{(5)}
+\bigg(\frac{3 \sqrt{15}}{14 c}\cW_3''+\frac{540 \sqrt{15}}{7c^2}\cL\cW_3-\frac{30}{c}\sqrt{\frac{15}{7}}\cW_5\bigg)\Wt_{-2}^{(5)}
+\bigg(\frac{45 \sqrt{15}}{7 c}\cW_5+\frac{10}{7 c}\sqrt{\frac{15}{7}}\cW_3''\nonumber\\
&+\frac{576}{7 c^2}\sqrt{\frac{15}{7}}\cL\cW_3\bigg)\Wt_{-2}^{(3)}
+\frac{45}{7 c}\sqrt{15}\cW_3\Wt_0^{(5)}+\bigg(\frac{216}{c^3}\cL^3+\frac{54}{5 c^2}\cL'^2+\frac{1500}{7 c^2}\cW_3^2+\frac{69}{5
c^2}\cL\cL''+\frac{1}{c}\sqrt{\frac{7}{15}}\cW_4''\nonumber\\
&+\frac{44}{c^2}\sqrt{\frac{21}{5}}\cL\cW_4+\frac{1}{20 c}\cL^{(4)} \bigg)\Wt_{-3}^{(4)}
-\frac{9}{7 c} \sqrt{15}\cW_3'\Wt_{-1}^{(5)} -\frac{6}{c}\cL'\Wt_0^{(4)}+\frac{18}{c}\cL\Wt_1^{(4)}-\frac{40}{7 c}\sqrt{\frac{15}{7}}\cW_3'\Wt_{-1}^{(3)}\nonumber\\
&+\frac{120}{7 c}\sqrt{\frac{15}{7}}\cW_3\Wt_0^{(3)}\bigg)
-\mu_4'
\bigg(\Wt_2^{(4)}+\frac{16}{c}\cL\Wt_0^{(4)}
-\bigg(\frac{11 \sqrt{15}}{392 c}\cW_3^{(3)}+\frac{507 \sqrt{15}}{49 c^2}\cL\cW_3'+\frac{759 \sqrt{15}}{49 c^2}\cW_3\cL'\nonumber\\
&-\frac{75}{28 c}\sqrt{\frac{15}{7}}\cW_5'\bigg)\Wt_{-4}^{(5)}
+\bigg(\frac{17}{60 c}\cL^{(3)}+\frac{241}{5 c^2}\cL\cL'+\frac{4}{c}\sqrt{\frac{7}{15}}\cW_4'\bigg)\Wt_{-3}^{(4)}
+\bigg(-\frac{19 \sqrt{15}}{98 c}\cW_3''-\frac{2682 \sqrt{15}}{49 c^2}\cL\cW_3\nonumber\\
&+\frac{120}{7 c}\sqrt{\frac{15}{7}}\cW_5\bigg)\Wt_{-3}^{(5)}-\frac{80}{7 c}\sqrt{\frac{15}{7}}\cW_3\Wt_{-1}^{(3)}-\frac{39 \sqrt{15} \cW_3\Wt_{-1}^{(5)}}{7 c}+\frac{11 \cL'\Wt_{-1}^{(4)}}{2 c}-\bigg(\frac{396}{5 c^2}\cL^2+\frac{6}{c}\sqrt{\frac{21}{5}}\cW_4\nonumber\\
&+\frac{7}{5 c} \cL'' \bigg)\Wt_{-2}^{(4)}+\frac{8 \sqrt{15}}{7c}\cW_3'\Wt_{-2}^{(5)}+\frac{30}{7 c}\sqrt{\frac{15}{7}}\cW_3'\Wt_{-2}^{(3)}\bigg)
+\mu_4''\bigg(\frac{1}{2}\Wt_1^{(4)}+\bigg(\frac{136}{5 c^2}\cL^2+\frac{13}{20 c}\cL''+\frac{1}{c}\sqrt{\frac{35}{3}}\cW_4\bigg)\Wt_{-3}^{(4)}\nonumber\\
&+\bigg(\frac{17 \sqrt{15}}{196 c}\cW_3''+\frac{876\sqrt{15}}{49 c^2}\cL\cW_3-\frac{123}{28 c}\sqrt{\frac{15}{7}}\cW_5\bigg)\Wt_{-4}^{(5)}-\frac{5}{2 c}\cL'\Wt_{-2}^{(4)}+\frac{7}{c}\cL\Wt_{-1}^{(4)}-\frac{\sqrt{15}}{2 c}\cW_3'\Wt_{-3}^{(5)}\nonumber\\
&+\frac{33}{14 c}\sqrt{15}\cW_3\Wt_{-2}^{(5)}+\frac{26}{7 c}\sqrt{\frac{15}{7}}\cW_3\Wt_{-2}^{(3)}\bigg)
+\mu _4^{(3)}\bigg(-\frac{1}{6} \Wt_0^{(4)}+\frac{3}{4c}\cL'\Wt_{-3}^{(4)}-\frac{2}{c}\cL\Wt_{-2}^{(4)}+\frac{\sqrt{15}}{7 c}\cW_3'\Wt_{-4}^{(5)}\nonumber
 \end{align}
 \begin{align}
&-\frac{9 \sqrt{15}}{14c}\cW_3\Wt_{-3}^{(5)}\bigg)
+\mu _4^{(4)}\bigg(\frac{1}{24}\Wt_{-1}^{(4)}+\frac{5}{12 c}\cL\Wt_{-3}^{(4)}+\frac{\sqrt{15}}{8 c}\cW_3\Wt_{-4}^{(5)}\bigg)
-\frac{\mu_4^{(5)}}{120}\Wt_{-2}^{(4)}+\frac{\mu _4^{(6)}}{720}\Wt_{-3}^{(4)}
 \end{align}
 \begin{align}
&a_{t}^{( \mu_{5})}=
 \mu _5\Bigg(\Wt_4^{(5)}+\frac{24}{c}\mathcal{W}_5\Lt_{-1} \nonumber\\
&+\bigg(\frac{216}{c^2}\mathcal{L}^2+\frac{2}{c}\mathcal{L}''-\frac{4 \sqrt{105}}{c}\mathcal{W}_4\bigg)\Wt_0^{(5)}+\bigg(\frac{1440}{7 \sqrt{7}
   c^2}\mathcal{W}_3^2+\frac{240 \sqrt{15}}{7 c^2}\mathcal{L}\mathcal{W}_4+\frac{2}{c}\sqrt{\frac{5}{3}}\mathcal{W}_4''\bigg)\Wt_{-2}^{(3)}\nonumber\\
&+\bigg(-\frac{2}{5}\mathcal{L}^{(3)}-\frac{672}{5 c^2}\mathcal{L}\mathcal{L}'+\frac{4}{c}\sqrt{\frac{21}{5}}\mathcal{W}_4'\bigg)\Wt_{-1}^{(5)}
   +\bigg(-\frac{2}{\sqrt{15} c}\mathcal{W}_3^{(3)}+\frac{4}{c}\sqrt{\frac{21}{5}}\mathcal{W}_5'-\frac{736}{7 c^2}\sqrt{\frac{3}{5}}\mathcal{W}_3\mathcal{L}'\nonumber\\
  &-\frac{912}{7 c^2}\sqrt{\frac{3}{5}}\mathcal{L}\mathcal{W}_3'\bigg)\Wt_{-2}^{(4)}
   +\bigg(-\frac{1}{105 c}\mathcal{L}^{(5)}+\frac{2}{\sqrt{105} c}\mathcal{W}_4^{(3)}-\frac{25056}{35
   c^3}\mathcal{L}^2\mathcal{L}'-\frac{188}{35 c^2}\mathcal{L}\mathcal{L}^{(3)}-\frac{376}{35 c^2}\mathcal{L}'\mathcal{L}''\nonumber\\
  &-\frac{1200}{7  c^2}\mathcal{W}_3\mathcal{W}_3'+\frac{432}{7 c^2}\sqrt{\frac{15}{7}}\mathcal{W}_4\mathcal{L}'+\frac{1864}{7
   c^2}\sqrt{\frac{3}{35}}\mathcal{L}\mathcal{W}_4'\bigg)\Wt_{-3}^{(5)}+\bigg(\frac{90000}{49 c^3}\mathcal{L} \mathcal{W}_3^2+\frac{1296}{c^4}\mathcal{L}^4+\frac{47}{35
   c^2}\mathcal{L}''^2\nonumber\\
  &+\frac{150}{7 c^2}\mathcal{W}_3'^2+\frac{420}{c^2}\mathcal{W}_4^2+\frac{1}{840 c}\mathcal{L}^{(6)}-\frac{1}{4 \sqrt{105}
   c}\mathcal{W}_4^{(4)}+\frac{4728}{35 c^3}\mathcal{L}^2\mathcal{L}''+\frac{1488}{7 c^3}\mathcal{L}\mathcal{L}'^2+\frac{27}{35
   c^2}\mathcal{L}\mathcal{L}^{(4)}\nonumber\\
  &+\frac{141}{70 c^2}\mathcal{L}'\mathcal{L}^{(3)}+\frac{225}{7 c^2}\mathcal{W}_3\mathcal{W}_3''-\frac{990}{\sqrt{7}
   c^2}\mathcal{W}_3\mathcal{W}_5-\frac{624}{c^3}\sqrt{\frac{15}{7}}\mathcal{L}^2\mathcal{W}_4-\frac{68}{7
   c^2}\sqrt{\frac{15}{7}}\mathcal{W}_4\mathcal{L}''-\frac{282}{7 c^2}\sqrt{\frac{3}{35}}\mathcal{L}\mathcal{W}_4''\nonumber\\
  & -\frac{503}{7
   c^2}\sqrt{\frac{3}{35}}\mathcal{L}'\mathcal{W}_4'\bigg)\Wt_{-4}^{(5)}+\bigg(\frac{1}{3 \sqrt{15} c}\mathcal{W}_3^{(4)}+\frac{1440 \sqrt{15}}{7
   c^3}\mathcal{L}^2\mathcal{W}_3+\frac{596}{7 \sqrt{15} c^2}\mathcal{L}\mathcal{W}_3''+\frac{824}{7 \sqrt{15} c^2}\mathcal{L}'\mathcal{W}_3'\nonumber\\
  &-\frac{2}{c}\sqrt{\frac{7}{15}}\mathcal{W}_5''+\frac{136}{7
   c^2}\sqrt{\frac{3}{5}}\mathcal{W}_3\mathcal{L}''-\frac{104}{c^2}\sqrt{\frac{15}{7}}\mathcal{L}\mathcal{W}_5+\frac{112 \sqrt{7}}{c^2}\mathcal{W}_3
   \mathcal{W}_4\bigg)\Wt_{-3}^{(4)}+\bigg(\frac{864}{c^3}\mathcal{L}^3+\frac{112}{5 c^2}\mathcal{L}'^2\nonumber\\
  &+\frac{2400}{7 c^2}\mathcal{W}_3^2+\frac{1}{15 c}\mathcal{L}^{(4)}-\frac{2}{c}\sqrt{\frac{7}{15}}\mathcal{W}_4''-\frac{272}{c^2}\sqrt{\frac{15}{7}}\mathcal{L} \mathcal{W}_4+\frac{152}{5 c^2}\mathcal{L}
   \mathcal{L}''\bigg)\Wt_{-2}^{(5)}+\bigg(-\frac{4 \sqrt{105}}{c}\mathcal{W}_5\nonumber\\
  &+\frac{576 \sqrt{15}}{7 c^2}\mathcal{L}\mathcal{W}_3+\frac{2}{c}\sqrt{\frac{5}{3}} \mathcal{W}_3''\bigg)\Wt_{-1}^{(4)}-\frac{8}{c}\mathcal{L}'\Wt_1^{(5)}+\frac{24}{c}\mathcal{L}\Wt_2^{(5)}
   +\frac{8 \sqrt{15}}{c}\mathcal{W}_3\Wt_1^{(4)}+\frac{8\sqrt{15}}{c}\mathcal{W}_4\Wt_0^{(3)}\nonumber\\
  &-\frac{8}{c}\sqrt{\frac{5}{3}}\mathcal{W}_3'\Wt_0^{(4)}-\frac{8}{c}\sqrt{\frac{5}{3}}\mathcal{W}_4'\Wt_{-1}^{(3)}\Bigg)
    +\mu_5'\bigg(-\Wt_3^{(5)}+\bigg(-\frac{876}{5 c^2}\mathcal{L}^2-\frac{19}{10 c}\mathcal{L}''+\frac{14}{c}\sqrt{\frac{21}{5}}\mathcal{W}_4\bigg)\Wt_{-1}^{(5)}\nonumber\\
  &+\bigg(\frac{7}{4 \sqrt{15} c}\mathcal{W}_3^{(3)}-\frac{7}{c}\sqrt{\frac{7}{15}}\mathcal{W}_5'+\frac{468}{7 c^2}\sqrt{\frac{3}{5}}\mathcal{W}_3\mathcal{L}'+\frac{599}{7 c^2}\sqrt{\frac{3}{5}}\mathcal{L}\mathcal{W}_3'\bigg)\Wt_{-3}^{(4)}+\bigg(\frac{31}{3360 c}\mathcal{L}^{(5)}-\frac{13}{8 \sqrt{105}
   c}\mathcal{W}_4^{(3)}\nonumber\\
  &+\frac{16482}{35 c^3}\mathcal{L}^2\mathcal{L}'+\frac{613}{140 c^2}\mathcal{L}\mathcal{L}^{(3)}+\frac{517}{56
   c^2}\mathcal{L}'\mathcal{L}''+\frac{3075}{28 c^2}\mathcal{W}_3\mathcal{W}_3'-\frac{1119}{7 c^2}\sqrt{\frac{3}{35}}\mathcal{L}\mathcal{W}_4'-\frac{1198}{7
   c^2}\sqrt{\frac{3}{35}}\mathcal{W}_4\mathcal{L}'\bigg)\Wt_{-4}^{(5)}\nonumber\\
   &+\bigg(-\frac{20088}{35 c^3}\mathcal{L}^3-\frac{666}{35 c^2}\mathcal{L}'^2-\frac{1500}{7 c^2}\mathcal{W}_3^2-\frac{9}{140 c}\mathcal{L}^{(4)}+\frac{11}{\sqrt{105} c}\mathcal{W}_4''-\frac{877}{35
   c^2}\mathcal{L}\mathcal{L}''+\frac{5324}{7 c^2}\sqrt{\frac{3}{35}}\mathcal{L}\mathcal{W}_4\bigg)\Wt_{-3}^{(5)}\nonumber\\
 &+\bigg(-\frac{17}{2 \sqrt{15} c}\mathcal{W}_3''+\frac{2 \sqrt{105}}{c}\mathcal{W}_5-\frac{1772}{7 c^2}\sqrt{\frac{3}{5}}\mathcal{L} \mathcal{W}_3\bigg)\Wt_{-2}^{(4)}
 -\frac{14}{c} \sqrt{\frac{5}{3}}\mathcal{W}_4\Wt_{-1}^{(3)} +\frac{15 \mathcal{L}'}{2 c}\Wt_0^{(5)}+\bigg(\frac{23}{60 c}\mathcal{L}^{(3)}\nonumber\\
 &-\frac{3}{c}\sqrt{\frac{21}{5}}\mathcal{W}_4'+\frac{554}{5 c^2}\mathcal{L} \mathcal{L}'\bigg)\Wt_{-2}^{(5)}-\frac{22}{c}\mathcal{L}\Wt_1^{(5)}-\frac{6 \sqrt{15}}{c}\mathcal{W}_3\Wt_0^{(4)}+\frac{11}{2
   c}\sqrt{\frac{5}{3}}\mathcal{W}_4'\Wt_{-2}^{(3)}+\frac{13}{2 c}\sqrt{\frac{5}{3}}\mathcal{W}_3'\Wt_{-1}^{(4)}
   \bigg)\nonumber\\
&+\mu_5''\bigg(\frac{1}{2}\Wt_2^{(5)}+\bigg(-\frac{11}{60 c}\mathcal{L}^{(3)}-\frac{223}{5 c^2}\mathcal{L}\mathcal{L}'+\frac{54}{7
   c}\sqrt{\frac{3}{35}}\mathcal{W}_4'\bigg)\Wt_{-3}^{(5)}+\bigg(\frac{346}{5 c^2}\mathcal{L}^2+\frac{9}{10
   c}\mathcal{L}''-\frac{33}{c}\sqrt{\frac{3}{35}}\mathcal{W}_4\bigg)\Wt_{-2}^{(5)}\nonumber\\
&+\bigg(\frac{101}{28 \sqrt{15}
   c}\mathcal{W}_3''-\frac{10}{c}\sqrt{\frac{5}{21}}\mathcal{W}_5+\frac{517}{7
   c^2}\sqrt{\frac{3}{5}}\mathcal{L}\mathcal{W}_3\bigg)\Wt_{-3}^{(4)}+\bigg(\frac{6144}{35 c^3}\mathcal{L}^3+\frac{2227}{280
   c^2}\mathcal{L}'^2+\frac{12225}{196 c^2}\mathcal{W}_3^2\nonumber\\
&+\frac{13}{420 c}\mathcal{L}^{(4)}-\frac{239}{56 \sqrt{105} c}\mathcal{W}_4''+\frac{352}{35
   c^2}\mathcal{L}\mathcal{L}''-\frac{1362}{7 c^2}\sqrt{\frac{3}{35}}\mathcal{L}\mathcal{W}_4\bigg)\Wt_{-4}^{(5)}-\frac{7 \mathcal{L}'}{2
   c}\Wt_{-1}^{(5)}-\frac{92 \mathcal{W}_3'}{7 \sqrt{15} c}\Wt_{-2}^{(4)}\nonumber
 \end{align}
\begin{align}
&+\frac{10}{c}\mathcal{L}\Wt_0^{(5)}+\frac{31 \sqrt{15}}{14 c}\mathcal{W}_3\Wt_{-1}^{(4)}+\frac{29}{7
   c}\sqrt{\frac{5}{3}}\mathcal{W}_4\Wt_{-2}^{(3)}
   \bigg)+\mu _5^{(3)}\bigg(-\frac{1}{6}\Wt_1^{(5)}-\frac{3 \mathcal{L}}{c}\Wt_{-1}^{(5)}+\bigg(-\frac{88}{5
   c^2}\mathcal{L}^2-\frac{17}{60 c}\mathcal{L}''\nonumber\\
&+\frac{149}{7 \sqrt{105} c}\mathcal{W}_4\bigg)\Wt_{-3}^{(5)}+\bigg(\frac{7}{120 c}\mathcal{L}^{(3)}-\frac{311}{56 \sqrt{105}
   c}\mathcal{W}_4'+\frac{58}{5 c^2}\mathcal{L}\mathcal{L}'\bigg)\Wt_{-4}^{(5)}-\frac{113 \mathcal{W}_3}{14 \sqrt{15} c}\Wt_{-2}^{(4)}+\frac{13 \mathcal{L}'}{12
   c}\Wt_{-2}^{(5)}\nonumber\\
&+\frac{33}{28 c}\sqrt{\frac{3}{5}}\mathcal{W}_3'\Wt_{-3}^{(4)}
   \bigg)
+\mu _5^{(4)}\bigg(\frac{1}{24}\Wt_0^{(5)}+\frac{2 \mathcal{L}}{3
   c}\Wt_{-2}^{(5)}+\bigg(\frac{16}{5 c^2}\mathcal{L}^2+\frac{1}{15 c}\mathcal{L}''-\frac{23}{7 \sqrt{105} c}\mathcal{W}_4\bigg)\Wt_{-4}^{(5)}-\frac{1}{4
   c}\mathcal{L}'\Wt_{-3}^{(5)}\nonumber\\
&+\frac{41}{28 \sqrt{15} c}\mathcal{W}_3\Wt_{-3}^{(4)}
   \bigg)
+\mu _5^{(5)}\bigg(-\frac{1}{120}\Wt_{-1}^{(5)}-\frac{7 \mathcal{L}}{60
   c}\Wt_{-3}^{(5)}+\frac{11 \mathcal{L}'}{240 c}\Wt_{-4}^{(5)}
   \bigg)
+\mu _5^{(6)}\bigg(\frac{1}{720}\Wt_{-2}^{(5)}+\frac{\mathcal{L}}{60 c}\Wt_{-4}^{(5)}\bigg)\nonumber\\
&-\frac{\mu_5^{(7)}}{5040}\Wt_{-3}^{(5)}+\frac{\mu _5^{(8)}}{40320}\Wt_{-4}^{(5)}
 \end{align}
\end{subequations}




\end{document}